\documentclass[10pt, journal, compsoc]{IEEEtran}

\usepackage{algorithm}
\usepackage{algorithmic}
\usepackage{amsmath,amssymb,amsfonts}
\usepackage{booktabs}
\usepackage{cite}
\usepackage{color}
\usepackage{dsfont}
\usepackage{enumerate}
\usepackage{longtable}
\usepackage{multirow}
\usepackage{graphicx}
\usepackage{setspace}
\usepackage{subfigure}
\usepackage{textcomp}
\usepackage{url}
\usepackage[breaklinks]{hyperref}
\usepackage{breakurl}
\usepackage{threeparttable}

\allowdisplaybreaks[2]

\newtheorem{lemma}{Lemma}

\newtheorem{theorem}{Theorem}

\begin{document}

\title{
POTUS: Predictive Online Tuple Scheduling for
Data Stream Processing Systems
}

\author{Xi Huang,~\IEEEmembership{Student Member,~IEEE,} 
Ziyu Shao$^*$,~\IEEEmembership{Senior Member,~IEEE,}
Yang Yang,~\IEEEmembership{Fellow,~IEEE}%
\interfootnotelinepenalty=100000
\thanks{
$^*$ The corresponding author of this work is Ziyu Shao. 

X. Huang, Z. Shao, and Y. Yang are with the School of Information Science and Technology, ShanghaiTech University, Shanghai, China. (E-mail: \{huangxi, shaozy, yangyang\}@shanghaitech.edu.cn)}
}
 
\IEEEtitleabstractindextext{%
\begin{abstract}

Most online service providers deploy their own data stream processing systems in the cloud to conduct large-scale and real-time data analytics. 
However, such systems, e.g., Apache Heron, often adopt naive scheduling schemes to distribute data streams (in the units of tuples) among processing instances, which may result in workload imbalance and system disruption. 
Hence, there still exists a mismatch between the temporal variations of data streams and such inflexible scheduling scheme designs. 
Besides, the fundamental benefits of predictive scheduling to data stream processing systems also remain unexplored.
In this paper, we focus on the problem of tuple scheduling with predictive service in Apache Heron. 
With a careful choice in the granularity of system modeling and decision making, we formulate the problem as a stochastic network optimization problem and propose \textit{POTUS}, an online predictive scheduling scheme that aims to minimize the response time of data stream processing by steering data streams in a distributed fashion. 
Theoretical analysis and simulation results show that POTUS achieves an ultra-low response time with stability guarantee.
Moreover, POTUS only requires mild-value of future information to effectively reduce the response time, even with mis-prediction.  

\end{abstract}

}

\maketitle

\IEEEraisesectionheading{\section{Introduction}\label{sec:introduction}}
\IEEEPARstart{T}{o} facilitate online services for ever-increasing and diversified user demands, more and more applications are deployed onto the data stream processing systems hosted in the cloud to conduct large-scale and real-time data analytics. 
For example, Twitter's backend applications are hosted in their own data centers to process billions of continuously-generated events per day from all over the world\cite{fu2017twitter}. 

Among the most updated data stream processing systems\cite{s4,samza,storm,kulkarni2015twitter}, Apache Heron\cite{kulkarni2015twitter} stands out with its highly modularized design and enjoys a wide adoption by numerous organizations such as Twitter and Facebook.
Within Apache Heron, the processing procedure of each application is described as a \textit{directed acyclic graph} (DAG).
In the graph, each vertex denotes a processing component with particular operations and each directed edge corresponds to the data stream flows between successive components \cite{kulkarni2015twitter}.
Upon deployment, each component is initialized as multiple independent processes, \textit{a.k.a.} \textit{Heron instances}, which are assigned to containers in different servers to ensure performance isolation and flexibility of resource management.
To avoid the high costs induced by instance migration and rescaling, the placement of instances is usually adjusted infrequently and hence can be deemed fixed at the time scale of data stream processing. 
At runtime, data streams are discretized as \textit{tuples}. 
Each tuple contains a list of values that are output by instances of the previous component and to be the input of the next component's instances.

\begin{table*}[!t]
{
\centering
\caption{Comparison between our work and the most related works}
\label{table: related works}
\renewcommand\arraystretch{1}
\setlength{\tabcolsep}{1.2mm}{
\begin{threeparttable}
\begin{tabular}{|c|c|c|c|c|c|c|}
\hline
\multirow{2}{*}{}
& Scheduling 
& Optimization
& Queue Stability
& Predictive
& General DAG
& Target
\\
	& Type
	& Metrics
	& Guarantee
	& Scheduling
	& Topology 
	& System Type
	\\ 
\hline
\cite{nasir2015power} & Tuple scheduling & Load imbalance &  &  & $\bullet$ & General DSP
\\
\hline
\multirow{2}{*}{\cite{rivetti2015efficient}} 
& \multirow{2}{*}{Tuple scheduling} & Percentage of &  &  & \multirow{2}{*}{$\bullet$} & \multirow{2}{*}{General DSP} 
\\
& & load imbalance & & & &
\\
\hline
\multirow{2}{*}{\cite{rivetti2016online}} 
& \multirow{2}{*}{Tuple scheduling} & Average tuple &  &  & \multirow{2}{*}{$\bullet$} & \multirow{2}{*}{General DSP} 
\\
& & completion time & & & & 
\\
\hline
\multirow{2}{*}{\cite{caneill2016locality}} 
& \multirow{2}{*}{Tuple scheduling} & Total amounts of &  &  & \multirow{2}{*}{$\bullet$} & \multirow{2}{*}{General DSP} 
\\
& & cross-server traffic & & & & 
\\
\hline
\multirow{2}{*}{\cite{fang2017parallel}} 
& \multirow{2}{*}{Tuple scheduling} & Total migration cost \& &  &  & \multirow{2}{*}{$\bullet$} & \multirow{2}{*}{General DSP} 
\\
& & workload imbalance & & & & 
\\
\hline
\cite{destounis2016streaming}
& Tuple scheduling & Total throughput & $\bullet$ &  &  & General DSP 
\\
\hline
\cite{schneider2016dynamic}
& Tuple scheduling & Maximum blocking rate & $\bullet$ &  & $\bullet$ & General DSP 
\\
\hline
\multirow{3}{*}{\cite{zhou2019predictive}} 
& \multirow{3}{*}{Server Provisioning} & Costs of energy, server &  & \multirow{3}{*}{$\bullet$} &  & DSP for 
\\
& & switching \& outsourcing, & & & & edge-based
\\
& & bandwidth, delay & & & & IoT systems
\\
\hline
\textbf{Our Work} & Tuple scheduling & Communication costs & $\bullet$ & $\bullet$ & $\bullet$ & General DSP \\
\hline
\end{tabular}
\end{threeparttable}
\begin{tablenotes}
	\footnotesize
	 \item[*] $^*$ DSP stands for data stream processing.
\end{tablenotes}}
}
\end{table*}

For online services with real-time requirements, the way tuples are scheduled among Heron instances, \textit{a.k.a.} \textit{tuple scheduling}, 
has a significant impact on their quality-of-service. 
In fact, the temporal variations of tuple traffic dynamics\cite{dhalion}, 
if coped with improperly, may cause workload imbalance among instances -- with some of them being overloaded while the others staying idle, or even fatal system breakdown, 
leading to unexpectedly long response time to latency-sensitive applications\cite{t-storm}. 
In Apache Heron, by default, tuples produced by the instances of one component will be distributed uniformly at random to the instances of its next component. In the meantime, a naive back-pressure scheme is employed to throttle all ingress components 
once any instance in the system is found overloaded\cite{kulkarni2015twitter}.
Such a scheme, though easy to implement and responsive to traffic change, also brings about disadvantages. 
On one hand, it implicitly assumes the homogeneity of instances' processing capacity, which often times is not the case in practice \cite{fu2017twitter}.
In fact, the difference in processing capacities can still incur workload imbalance among instances. 
On the other hand, throttling-based back-pressure can severely hurt application performance with unexpected and undesirable tuple loss and congestion. 

By far, there still remains a mismatch between the constant temporal variations of data stream dynamics and the inflexible tuple scheduling scheme design. 
Faced with such a mismatch, it is an open problem to design an online and adaptive tuple scheduling scheme that achieves the workload balance among instances of different components.
Furthermore, the scheme should also be computationally efficient so that we can trade little decision-making overheads for a significant improvement in the system performance.
Moreover, inspired by the recent wide adoption of predictive scheduling in various different systems,\footnote{For example, Netflix prefetches videos of interest onto user-end devices based on user-behavior prediction\cite{NetflixPred}.} 
some natural questions come along: 
1) If tuple arrivals can be predicted ahead of a short time window, then what are the \textit{fundamental} benefits of such information to tuple scheduling? 
2) Considering that such predictive scheduling, if wrongly decided, may consume extra system resources, then what is the impact of mis-prediction on the scheduling? 
Answers to these questions are the crux to understand the endeavors that are worthy to put on predictive scheduling in data stream processing systems.

In this paper, by focusing on the tuple scheduling problem in Apache Heron, we address the above challenges and answer the questions. 
Our key results and contributions are summarized as follows. 
	
	\textbf{System Modeling and Formulation:} 
	We develop a novel system model with a careful choice in the granularities of both system state characterization and decision making. Particularly, our system model captures the interplay between successive instances at the granularity of tuples while characterizing the decision making on a per-time-slot basis to avoid the considerable overheads brought by per-tuple optimization.  
	Then we formulate the tuple scheduling problem with predictive service as a stochastic network optimization problem. 
	By exploiting the unique problem structure, we further transform the problem into a series of sub-problems over time slots. 

	\textbf{Algorithm Design:}
	Based on the problem transformation, we propose \textit{POTUS}, a predictive tuple scheduling scheme that steers tuples between successive processing instances in a distributed and online fashion.
	Our theoretical analysis shows that POTUS can achieve a near-optimal communication cost with queue stability guarantee.
	
	\textbf{Predictive Scheduling:}
	{To our best knowledge, this paper is the first to consider predictive tuple scheduling in general data stream processing systems}, which provides a new perspective in the field of data stream processing. 
	It also provides the first systematic study on the fundamental benefits of predictive scheduling and the impacts of mis-prediction on such systems.
	Our solution can also be adapted to other data stream processing systems \cite{samza,storm,kulkarni2015twitter}.
	
	\textbf{Experiment Verification and Investigation:} 
	Our simulation results show that, POTUS outperforms the state-of-the-art scheme in Heron by achieving a close-to-optimal communication cost and ultra-low response time. 
	In addition, we investigates the benefits of predictive scheduling. Notably, we find that only mild-value of future information suffices to significantly shorten the tuple response time, even in face of mis-prediction.
	
The rest of this paper is organized as follows. 
We first discuss related work in Section \ref{sec: related work}, then present our model and problem formulation in Section \ref{sec: problem formulation}. 
In Section \ref{sec: algorithm design}, we show the detailed design of POTUS, followed by its theoretical analysis. 
Section \ref{sec: simulation} presents simulation results and the corresponding analysis, while Section \ref{sec: conclusion} concludes this paper.

\section{Related Work}  \label{sec: related work}

{\noindent \textbf{Data stream processing (DSP) systems.}}
To date, the design of DSP systems has evolved over three generations\cite{kulkarni2015twitter} \cite{de2018distributed}. 
Since about $2009$, systems of the third-generation, \textit{e.g.}, \cite{samza}\cite{storm}\cite{flink, millwheel, spark_streaming}, have been proposed successively to accommodate large-scale and distributed cloud-based data stream processing with high scalability and resiliency guarantee.
Amongst such systems, Apache Storm \cite{storm} and its successor Apache Heron \cite{kulkarni2015twitter} have enjoyed the most intensive attraction from both industries and academia since released. 
Existing works mainly focus on optimizing the placement, re-scaling, and migration of processing instances to improve the average response time of tuple processing \cite{t-storm}\cite{drs, adaptive_on_off, stela, locality}, overall throughput\cite{dhalion}\cite{peng2015r}\cite{zhang2015nodes}, resource utilization\cite{huang2019mips}, and other performance metrics such as charges for communication \cite{gu2015general} and revenue loss due to QoS violation\cite{operator_rep_and_plc}\cite{poster}. 
The focuses of such works are orthogonal to and can be well integrated with our devised scheme.\footnote{
		In particular, given the provisioning, scaling, and placement of processing instances derived by the solutions in such works, our design can be initiated and directly applied to schedule tuples in real systems.}

{\noindent \textbf{Tuple scheduling.}} 
Regarding tuple scheduling, a common objective is to ensure the workload balance among instances and achieve ultra-low tuple response time. 
However, well-known systems such as Apache Storm generally provide no close-loop control to fulfill such goals. To this end, a number of works \cite{nasir2015power, rivetti2015efficient, rivetti2016online, caneill2016locality, fang2017parallel}\cite{schneider2016dynamic} have been proposed to improve the load balancing among processing instances by utilizing dedicated system dynamics, \textit{e.g.}, the locality of instances and the proportions of tuple content hashing at runtime. 
Such solutions, despite their effectiveness, generally require the instance migration across different servers or the accurate estimate about the statistics about the instance processing. 
These features may incur non-negligible overheads in practice.  
Meanwhile, to fix such issues, Apache Heron, the successor of Apache Storm, introduced a naive backpressure scheme, \textit{i.e.}, freezing the whole application once any instance is found overloaded. 
However, such a \textit{reactive} mechanism only handles workload imbalance when it occurs. 
{A better alternative is to schedule tuples between instances in a \textit{proactive} manner. 
To this end, previous work \cite{destounis2016streaming} introduced backpressure mechanism to conduct tuple scheduling in an online fashion and investigated the corresponding maximum achievable throughput. 
Nonetheless, their solution does not consider the communication costs (\textit{e.g.}, in terms of bandwidth consumption) that are incurred for tuple transmission. Moreover, predictive scheduling is not in the scope of their study. 
In comparison, our work systematically investigates the fundamental limits of benefits from predictive scheduling in such systems upon a joint optimization of communication costs and queue stability.
Our solution requires only limited information of system dynamics for the decision making with little overheads.}

{\noindent \textbf{Predictive scheduling.}}
Predictive scheduling has been applied to a wide range of computing systems \cite{huang2019sdn, huang2019nfv, gao2019fog}.
{Note that in the previous work \cite{huang2016backpressure}, the authors focus on the predictive scheduling problem for controlled queueing systems wherein multiple user queues are jointly served by a single server. Their model and results cannot be applied to data stream processing (DSP) systems. This is because in DSP systems, each application is usually formed as a directed acyclic graph (DAG) of processing instances and the instances are often distributed across different servers. The DAG nature of such applications implies that DSP systems are intrinsically more general queueing networks with more complicated queueing dynamics. Although their multi-queue-single-server model can be viewed as a special case of such queueing networks, it is non-trivial to extend their results to more general cases. 
Recently, Zhou \textit{et al.} \cite{zhou2019predictive} considered exploiting predictive information to optimize the activation of edge servers for data stream processing with a better energy efficiency. 
Different from such works, our study focuses on investigating the \textit{fundamental} limits of the benefits from predictive information for \textit{tuple scheduling} in such systems.
To our best knowledge, our work provides the first systematic investigation on such a problem with theoretical analysis and numerical evaluations.}

{\noindent \textbf{Traffic prediction in the cloud.}}
Prior works have proposed various methods for the traffic or workload prediction in various computing systems \cite{zhang2015nodes}\cite{khan2012workload, peng2014hadoopwatch, cortez2017resource, calheiros2014workload}. 
We note that our work is orthogonal to theirs, 
as our focus is on exploring the fundamental limits of benefits from predictive scheduling to data stream processing systems.

We summarize the comparison between our work and the most related works in Table \ref{table: related works}.

\section{Problem Formulation}  \label{sec: problem formulation}

\begin{figure}[t!]
	\centering
	\includegraphics[width=0.32\textwidth]{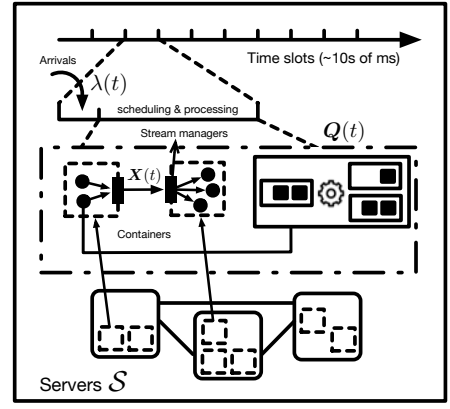}
	\caption{
		An overview of our system model. 
		Basically, the system proceeds over time slots and consists of a set of servers $S$, each hosting some containers. For data stream processing applications, their components' instances (solid circles) are packed into a number of containers with a fixed placement. Each instance maintains an input queue and several output queues to buffer the untreated and processed tuples (all denoted by solid squares) by the current instance, respectively.
		At the beginning of each time slot $t$, each instance receives a number $\lambda(t)$ of new tuples, serves some of them from its input queue, and outputs those processed to its output queues. 
		Then the stream manager in each container makes the scheduling decision $\boldsymbol{X}(t)$ to decide how the processed tuples are forwarded among the instances of successive components.
	}
	\label{motivation}
\end{figure}

We consider a Heron-based data stream processing system which hosts a number of applications over a cluster of 
servers. 
The system proceeds over time slots, each with a fixed length proportional to the average tuple processing latency \cite{storm}. 
During each time slot, stream managers that reside in the containers of servers jointly decide the scheduling of tuples among instances of successive components. 
An example of our system model is given in Figure \ref{motivation}.
In the following, we illustrate our model with our problem formulation. 
The key notations are summarized in Table \ref{notations}.

\begin{table}[!t]
	\caption{Key Notations} \label{notations}
	\centering
	\begin{tabular}{l|l}
	\toprule
	\hline
	Symbol & Description \\
	\hline
	$\mathcal{A}$ & Set of applications in the system \\
	\hline
	$\mathcal{C}$ & Set of components of all applications \\
	\hline
	$\mathcal{I}$ & Set of processing instances in the system\\
	\hline
	$\mathcal{K}$ & Set of all containers in the system\\
	\hline
	$\mathcal{I}_{C}(c)$ & Set of instances belonging to component $c$ \\
	\hline
	$\mathcal{I}_{K}(k)$ & Set of instances hosted in container $k$ \\
	\hline
	\multirow{2}*{$p(i)$} & Set of components that precede the \\
	& component of instance $i$ \\
	\hline
	\multirow{2}*{$n(i)$} & Set of components that are subsequent \\
	& to the component of instance $i$ \\
	\hline
	\multirow{2}*{$X_{i,i'}(t)$} & Variable that decides the number of tuples \\
	& to be sent from instance $i$ to $i'$ in time slot $t$ \\ 
	\hline
	$\gamma_{i}$ & Transmission capacity of instance $i$ \\
	\hline
	\multirow{2}*{$Q^{(\text{in})}_{i}(t)$} & Size of the input queue on instance $i$ at the \\
	& beginning of time slot $t$ \\
	\hline
	\multirow{3}*{$Q^{(\text{out})}_{i, c'}(t)$} & Size of the output queue on instance $i$ to the \\
	& instances of component $c'$ at the beginning \\
	& of time slot $t$ \\
	\hline
	\multirow{2}*{$\lambda_{i, c'}(t)$} & Number of new tuples generated on spout \\
	& instance $i$ to component $c'$ in time slot $t$ \\
	\hline
	\multirow{3}*{$U_{k, k'}(t)$} & The amounts of bandwidth consumed for \\
	& sending one tuple from container $k$ to \\
	& container $k'$ during time slot $t$ \\
	\hline
	$W_{i}$ & Size of the lookahead window on instance $i$ \\
	\hline
	\bottomrule
	\end{tabular}
\end{table}

\subsection{Streaming Application Model}

We define $\mathcal{A}$ as the set of applications running in the system.
Each application $a \in \mathcal{A}$ is denoted by a directed acyclic graph (DAG), \textit{a.k.a.} its \textit{topology}, with its processing components as nodes (denoted by set $\mathcal{C}_{a}$) and data stream flows as edges.\footnote{
In practice, the diameter of such application topology is usually not very large, mostly no more than three\cite{storm}. Nevertheless, our model is applicable to any data stream processing application with arbitrary DAG topology.}
Typically, in each topology, there are two types of components, \textit{i.e.}, \textit{spouts} and \textit{bolts}. 
On one hand, each spout loads and discretizes data streams into tuples, then forwards such tuples to its subsequent bolts. 
On the other hand, each bolt receives and processes tuples from its previous components (spouts or bolts), then sends the processed tuples to their successors.
Accordingly, we define $\mathcal{C} \triangleq \overset{.}{\bigcup}_{a \in \mathcal{A}}\mathcal{C}_{a}$ as the set of all components of applications in the system. Each component belongs to only one application topology.
Components with the same processing function in different applications are still deemed as distinct components in our model.

\subsection{Deployment Model}
We assume that all applications are deployed in a cluster of heterogeneous servers, denoted by set $\mathcal{S}$. 
{In particular, for scalability and reliability concerns, each component $c \in \mathcal{C}$ (either a spout or a bolt) is instantiated as multiple instances, denoted by set $\mathcal{I}_{C}(c)$.
Each instance of the same component has the same function.
Accordingly, we denote $\mathcal{I} \triangleq \overset{.}{\bigcup}_{c \in \mathcal{C}} \mathcal{I}_{C}(c)$ as the set of all processing instances in the system. 
For each instance $i$, we use $p(i)$ to denote the set of components that precede instance $i$'s component in its application topology. 
Meanwhile, we use $n(i)$ to denote the set of components that are subsequent to the component of instance $i$ in the topology.
Note that $p(i)=\emptyset$ for any spout instance $i$, while $n(i')=\emptyset$ for any terminal bolt instance $i'$.

Upon deployment, all instances in the system are packed into a set of containers in servers, denoted by set $\mathcal{K}$. Each instance runs as an independent process within a container.
Meanwhile, we assume that instances' placement is fixed over containers which is pre-determined by existing optimization schemes such as \cite{t-storm}\cite{huang2019mips}.\footnote{
Previous work \cite{zhou2019online} has shown that a joint optimization for instance provisioning, instance placement, and tuple scheduling can lead to a better performance. However, such joint optimization is usually viable only under particular assumptions. 
For general data stream processing systems, the joint scheme design remains non-trivial. 
Therefore, in our work, we take the first step to investigate the fundamental limits of benefits of predictive information for tuple scheduling given that the provisioning and placement of processing instances are pre-determined.}
Accordingly, for each container $k \in \mathcal{K}$, we denote the set of its hosted instances by $\mathcal{I}_{K}(k)$.}


\subsection{Tuple Scheduling Decision}
At runtime, within each time slot $t$, 
the stream manager in each container $k$ would make a set of tuple scheduling decisions $X_{i, *}(t)$ for each instance $i \in \mathcal{I}_{K}(k)$. 
Particularly, 
for $c' \in n(i)$ and $i' \in \mathcal{I}_{C}(c')$,
variable $X_{i, i'}(t)$ determines the number of tuples to send from instance $i$ to instance $i'$;
otherwise, $X_{i, i'}(t) = 0$.  
Note that, for each instance $i$, the maximum number of tuples that are allowed to transmit in one time slot is denoted by $\gamma_{i}$, \textit{i.e.},
\begin{equation}\label{constraint rate per instance}
	\setstretch{1.2}
	\sum_{c' \in n(i)}
	\sum_{i' \in \mathcal{I}_{C}(c')}
	X_{i, i'}(t) \le \gamma_{i}.
\end{equation}

\setlength{\textfloatsep}{3pt}

\subsection{Queueing Model}
Due to limited processing capacities, instances may not be able to serve all incoming tuples within one time slot. 
Instead, each instance maintains some queues to buffer tuples\cite{kulkarni2015twitter}. 
Particularly, each instance $i$ maintains at least one queue to buffer tuples that are either generated from data streams (if $i$ is a spout instance), or processed by the current instance (if $i$ is a bolt instance).
Depending on the topology, the buffered tuples may be sent to instances of more than one bolt. 
In our model, we distinguish the tuples to be sent to different targeted bolts.\footnote{
		We omit the output queues on the instances of terminal bolts.
	} 
Specifically, for each bolt $c \in n(i)$, we consider a virtual output queue on the current instance $i$ with a size of $Q^{(\text{out})}_{i, c'}(t)$ at the beginning of each time slot $t$, which buffers all tuples to be sent to the instances of bolt $c$.
Note that such a multi-queue model is logically equivalent to the physical output queue and captures the accurate tuple queueing dynamics during each time slot.

\begin{figure}[t!]
	\centering
	\includegraphics[width=0.44\textwidth]{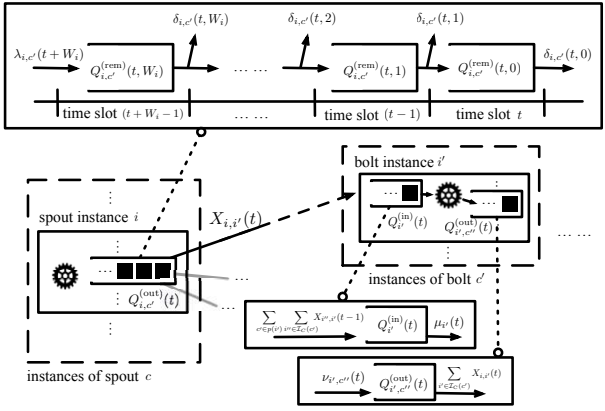}
	\caption{
		An illustration of the queueing dynamics on different instances.
	}
	\label{queueing_dynamics}
\end{figure}

If instance $i$ is a bolt instance, then besides the output queue, it also maintains an input queue to buffer the tuples sent from preceding instances.
We denote the size of its input queue at the beginning of each time slot $t$ by $Q^{(\text{in})}_{i}(t)$.
For simplicity, we define $\boldsymbol{Q}(t)$ as the set of all queue sizes at the beginning of each time slot $t$. 
Next, we elaborate the queueing dynamics in our model, as shown in Figure \ref{queueing_dynamics}.

\textbf{Queueing with Predictive Tuple Arrivals:}
{In the past few years, it has become viable to make online short-term predictions for data stream arrivals or user query traffic in various communication or computing systems (e.g., SDN, NFV, Fog). The viability of such practice mostly attributes to the rapid advancement of machine learning techniques during recent years. For data stream processing systems, by employing such prediction techniques, it is promising to conduct predictive scheduling by pre-admitting and pre-serving data streams for a better quality of service. For example, Netflix and Tik-Tok pre-schedule video clips onto user devices based on their user preference prediction\cite{NetflixPred}. Likewise, with real-time trend analysis, Twitter can proactively push users’ interested feeds and topics to promote the quality of user experience. However, prior to our work, the modeling and scheme design for the predictive scheduling of data streams, as well as the fundamental limits of its gains still remain unexplored. To this end, our work considers a data stream processing system in which future tuple arrivals in a limited lookahead time window can be perfectly predicted, pre-generated, and buffered at ingress processing instances (a.k.a. spout instances) in the system.}
Particularly, in our model, we do not assume any particular prediction techniques.\footnote{
In practice, prediction techniques usually vary in different degrees of trade-off between computational complexity and prediction accuracy. Particularly, the computational complexity of a given predictive technique is often used to characterize its prediction cost. Regarding the concrete justification of the costs and benefits of employing prediction for tuple scheduling, a further analysis with respect to particular prediction techniques is required.}
Instead, we consider the prediction results as the output from the other standalone modules with prediction techniques such as time-series prediction\cite{box2015time}. 
Accordingly, the predicted tuples can be pre-admitted by spout instances and acquire pre-service by the system.

\begin{figure}[t!]
	\centering
	\includegraphics[width=0.31\textwidth]{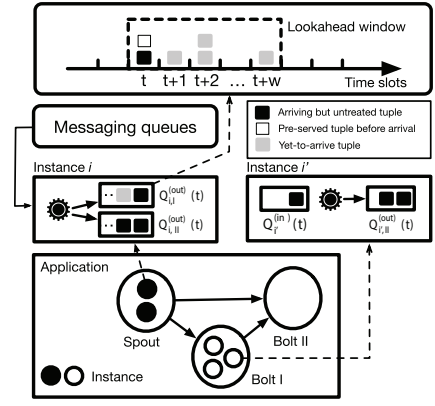}
	\caption{
		An illustration of the queueing model.
	}
	\label{queueing_model}
\end{figure}

Under such settings, the system workflow proceeds as follows. 
At the beginning of each time slot $t$, given instant system dynamics, stream managers make the tuple scheduling decisions $\boldsymbol{X}(t)$ for the instances in their respective containers. 
According to such decisions, each spout instance would produce new tuples (also including those pre-admitted ones);
in the meantime, both spout and bolt instances forward tuples from their output queues to their succeeding instances. 
Next, each bolt instance processes tuples from its input queue and generate new ones to output queues.
Such a queueing model is illustrated in Figure \ref{queueing_model}.

In the following, we specify the queueing dynamics in each time slot $t$ 
on spout and bolt instances, respectively.

\textbf{Queueing Dynamics on Spout Instances:}
At the beginning of each time slot $t$, 
for each spout instance $i$ and one of its successive components $c' \in n(i)$, 
a number of $\lambda_{i, c'}(t)$ ($\leq \lambda_{\text{max}}$ for some constant $\lambda_{\text{max}}$) new tuples would arrive at its output queue.
Such tuple arrivals are assumed \textit{i.i.d.} over time slots. 
Besides, instance $i$ is also assumed to have full access to the future tuple arrivals in a finite lookahead window of size $W_{i}$ ($\leq W_{\text{max}}$ for some constant
$W_{\text{max}}$), denoted by $\{ 
\lambda_{i, c'}(t + 1), \dots, \lambda_{i, c'}(t + W_{i})\}$.
The predicted tuples in the lookahead time window can be pre-generated and pre-served by the system. Hence, some tuples in the lookahead window may have been processed before. By defining $Q^{\text{(rem)}}_{i, c'}(t, w)$ ($0 \leq w \leq W_{i}$) as the number of untreated tuples in time $(t+w)$ by time slot $t$, we have
\begin{equation}
	0 \leq Q^{\text{(rem)}}_{i, c'}(t, w) \leq
	\lambda_{i, c'}(t+w).
\end{equation}
Note that $Q^{\text{(rem)}}_{i, c'}(t, 0)$ is 
the number of untreated tuples that actually arrive by time $t$.
Consequently, at the beginning of time slot $t$, the output queue for bolt $c' \in n(i)$ actually buffers all untreated tuples in the next $W_{i}$ time slots, with a total size of 
\begin{equation}
	Q^{\text{(out)}}_{i, c'}(t) = 
	\sum_{w=0}^{W_{i}} Q^{\text{(rem)}}_{i, c'}(t, w).
\end{equation}
We assume that with scheduling decision $\boldsymbol{X}(t)$, tuples are routed in a fully efficient manner \cite{huang2016backpressure}. 
That being said, by defining $\delta_{i, c'}(t, w)$ (for $0 \leq w \leq W_{i} - 1$) as the number of tuples that will be forwarded from queue $Q^{\text{(rem)}}_{i, c'}(t, w)$, 
to ensure that all tuples to be admitted by the time they arrive, we require that 
\begin{equation}
	\delta_{i, c'}(t, 0) = Q^{\text{(rem)}}_{i, c'}(t, w).
\end{equation}
We have the following update equations for $Q^{\text{(rem)}}_{i, c'}(t, w)$.
\begin{enumerate}
	\item[$\diamond$] For $0 \le w \le W_{i} - 1$,
	\begin{equation}
		Q^{\text{(rem)}}_{i, c'}(t+1, w) = 
		\left[ Q^{\text{(rem)}}_{i, c'}(t, w+1) - \delta_{i, c'}(t, w+1) \right]^{+};
	\end{equation}
	\item[$\diamond$] For $w = W_{i}$,
	\begin{equation}
		Q^{\text{(rem)}}_{i, c'}(t+1, W_{i}) = 
		\lambda_{i, c'}(t + W_{i} + 1),
	\end{equation}
\end{enumerate}
where $[x]^{+} \triangleq \max\{x, 0\}$. In other words, the total number of output tuples 
	$\sum_{i' \in \mathcal{I}_{C}(c')}\!\!X_{i, i'}(t)$ 
	from $Q^{\text{(out)}}_{i, c'}(t)$ in time slot $t$ is equal to
	$\sum_{w=0}^{W_i-1} \delta_{i, c'}(t, w)$.
Therefore, $Q^{\text{(out)}}_{i, c'}(t)$ is updated as follows,
\begin{equation}\label{qupdate-oqueue-on-spout}
	\begin{array}{l}
	Q^{\text{(out)}}_{i, c'}(t + 1) \\
	\displaystyle
	= 
	\bigg[  
		Q^{\text{(out)}}_{i, c'}(t) - 
		\!\!
		\sum_{i' \in \mathcal{I}_{C}(c')}
		\!\!
		X_{i, i'}(t)
	\bigg]^{+} \!\!\!\!\! +
	\lambda_{i, \eta}(t + W_{i} + 1).		
	\end{array}
\end{equation}

\textbf{Queueing Dynamics on Bolt Instances:}
At the beginning of each time slot $t$, each bolt instance $i$ has an input queue of size $Q^{(\text{in})}_{i}(t)$ with a processing capacity of ${\mu}_{i}(t)$ ($\leq \mu_{\text{max}}$ for some constant $\mu_{\text{max}}$).
After processing, instance $i$ will generate $\nu_{i, c'}(t)$ tuples to its output queue $Q^{(\text{out})}_{i, c'}(t)$, $\forall\, c' \in n(i)$. 
All such information is attainable from system modules such as the metric managers and stream managers at runtime\cite{kulkarni2015twitter}.
Then, with scheduling decision $\boldsymbol{X}(t)$, 
$Q^{\text{(in)}}_{i}(t)$ and $Q^{\text{(out)}}_{i, c'}(t)$ are updated as follows, respectively,
\begin{equation}\label{qupdate-iqueue-on-bolt}
	\begin{array}{l}
	\!\!\!\!\!\!\!\!\!Q^{\text{(in)}}_{i}(t + 1) 
	\\
	\displaystyle
	=
	\bigg[
		Q^{\text{(in)}}_{i}(t) + 
		\!\!
		\sum_{c' \in p(i)}
		\sum_{i' \in \mathcal{I}_{C}(c')}\!\!\!\!
		X_{i', i}(t-1) 
		- \mu_{i}(t)
	\bigg]^{+}\!\!.		
	\end{array}
\end{equation}
\begin{equation}\label{qupdate-oqueue-on-bolt}
	Q^{\text{(out)}}_{i, c'}(t + 1) 
	=
	\bigg[
		Q^{\text{(out)}}_{i, c'}(t)
		-
		\sum_{i' \in \mathcal{I}_{C}(c')}
		\!\!
		X_{i, i'}(t)
	\bigg]^{+} \!\!\!
	+ \nu_{i, c'}(t).
\end{equation}

In addition, for each instance $i \in \mathcal{I}$ and each component $c' \in n(i)$, we aim to maximize the utilization of the transmission capacity by restricting that
\begin{equation}\label{constraint on max-rate}
	\setstretch{1.2}
	\sum_{i' \in \mathcal{I}_{\text{C}}(c')}
		X_{i, i'}(t)
	\leq	
	Q^{\text{(out)}}_{i, c'}(t).
\end{equation}

\subsection{Optimization Objectives:}

\textbf{Communication Cost:}
{To enable responsive online services, it is essential to minimize the \textit{communication costs} in data stream processing systems. 
In practice, such costs can be characterized by the bandwidth consumption for sending tuples across containers and servers. 
In practice, excessive bandwidth consumptions are undesirable and may even lead to prohibitive charges when data streams are transmitted between instances across different areas (\textit{e.g.}, data centers \cite{gu2015general} or edge nodes \cite{zhou2019predictive}. In our model, we assume that such communication costs are proportional to the number of forwarded tuples.
Particularly, we define $U_{k, k'}(t)$ as the amounts of bandwidth consumed for sending a tuple from container $k$ to container $k'$ during each time slot $t$. Then we use $\boldsymbol{U}(t)$ to denote the set of all such costs. 
Note that $\boldsymbol{U}(t)$ is known \textit{a priori} to the system upon decision making. 
Then given decision $\boldsymbol{X}(t)$, the total communication costs are given by}
\begin{equation}
	\begin{array}{c}
		\displaystyle
		\Theta(t) \triangleq 
		\hat{\Theta}(\boldsymbol{X}(t)) 
		=
		\sum_{i,\, i'  \in \mathcal{I}} 
		X_{i, i'}(t)
		{U}_{k(i), k(i')}(t),
	\end{array}
\end{equation}
where $k(i)$ and $k(i')$ denote the containers that host instances $i$ and $i'$, respectively.

\textbf{Queueing Stability:} 
Reducing the queueing delay also conduces to shortening tuple response time.
By \textit{Little's Theorem}\cite{little1961proof}, the queue size is proportional to the average response time.  
Besides, it has also been verified that overloading any instance will cancel the benefit from reducing communication costs \cite{t-storm}. 
Therefore, we should not only forward as many tuples as possible but also balance the queues on different instances. 
We denote the weighted total queue backlog size in time slot $t$ as 
\begin{equation}\label{def_total_q}
	\setstretch{1.4}
	h(t) \triangleq
	\hat{h}(\boldsymbol{Q}(t)) = 
	\sum_{i \in \mathcal{I}} 
	Q^{(\text{in})}_{i}(t)	
	+ \beta
	\sum_{i \in \mathcal{I}}
	\sum_{c' \in n(i)}
	Q_{i, c'}^{\text{(out)}}(t),
\end{equation}
where $\beta$ is a positive constant that 
weighs the importance of stabilizing the output queues compared to input queues on instances.
In practice, one can set the value of parameter $\beta$ proportional to the ratio between the capacity on the input queue backlogs and the output queue backlogs of instances. 
Accordingly, we define queueing stability \cite{neely2010stochastic} as 
\begin{equation}\label{constraint queue stability}
	\setstretch{1.6}
	\begin{array}{c}
	\displaystyle
	\limsup_{T \to \infty}
	\frac{1}{T}
	\sum_{t=0}^{T-1} \mathbb{E}\left\{ h(t) \right\} < \infty,		
	\end{array}
\end{equation}
which ensures that the total number of tuples accumulating in the system will not grow to infinity as time goes by, so that no instances will be overloaded and each tuple can receive a timely processing.

\subsection{Problem Formulation}
Based on the above model, we formulate the {tuple scheduling problem} as the following stochastic network optimization problem with the aim to minimize the long-term time-averaged communication costs with stability guarantee.
\begin{equation}\label{problem-def-small-time-scale}
	\begin{array}{cl}
		\underset{\{\boldsymbol{X}(t)\}_{t}}{\text{Minimize}}
		& 
		\displaystyle
		\limsup_{T \to \infty}
		\frac{1}{T} \sum_{t = 0}^{T - 1}
		\mathbb{E}\left\{  \hat{\Theta}\left( \boldsymbol{X}(t) \right) \right\}
		\\
		\text{Subject to} 
		& 
		(\ref{constraint rate per instance}),
		(\ref{constraint on max-rate}),
		(\ref{constraint queue stability}).
	\end{array}
\end{equation}
In other words, we seek to find a tuple scheduling scheme which conducts a series of decision making across time slots to ensure that all generated tuples will be processed in a timely fashion with low communication costs in the long run;
meanwhile, no instances will ever be overloaded. Such a problem is in general hard to be solved directly because of the temporal variations of system dynamics, the online nature of tuple arrivals, and the complicated interactions among instances with resource constraints (\ref{constraint rate per instance}) and (\ref{constraint on max-rate}).
Besides, the involvement of predicted information further complicates the problem.

\section{Algorithm Design and Analysis}  \label{sec: algorithm design}
In this section, we conduct a non-trivial transformation on problem (\ref{problem-def-small-time-scale}), based on which we propose \textit{POTUS}, a Predictive Online Tuple Scheduling scheme that solves problem (\ref{problem-def-small-time-scale}) approximately optimally in a distributed fashion. 
In the following subsections, we first show how we transform the original problem, followed by the detailed design of POTUS and the corresponding theoretical analysis.  

\subsection{Algorithm Design}
Instead of solving problem (\ref{problem-def-small-time-scale}) directly, we adopt Lyapunov optimization techniques \cite{neely2010stochastic} to transform the long-term stochastic optimization problem (\ref{problem-def-small-time-scale}) into a series of sub-problems over time slots, which is specified by the following lemma.
\begin{lemma}
	\textit{
		By applying Lyapunov optimization techniques and the concept of opportunistically minimizing an expectation, problem (\ref{problem-def-small-time-scale}) can be transformed into the following problem to be solved during each time slot $t$:
		\begin{equation}\label{subpr-for-tuple-routing}
	\begin{array}{cl}
		\underset{
			\boldsymbol{X}
		}{\text{Minimize}}
		& \displaystyle
		\sum_{i \in \mathcal{I}}
		\sum_{c' \in n(i)}
		\sum_{i' \in \mathcal{I}_{\text{C}}(c')}
		l_{i, i'}(t)
		X_{i, i'} \\
		\text{Subject to} &
		\displaystyle
		\sum_{c' \in n(i)}
		\sum_{i' \in \mathcal{C}(c')}
		X_{i, i'}(t) \le \gamma_{i},\
		\forall\, i \in \mathcal{I}, \\
		& \displaystyle
		\sum_{i' \in \mathcal{I}_{\text{C}}(c')} 
			X_{i, i'} 
		\le Q_{i, c'}^{\text{(out)}}(t) \  
		\forall\ i \text{ and } c' \in n(i),
	\end{array}
\end{equation}
where we define
\begin{equation}\label{l-def}
	l_{i, i'}(t) \triangleq 		
		V \cdot U_{c(i), c(i')}(t)
		+ Q^{\text{(in)}}_{i'}(t)
		- \beta Q^{\text{(out)}}_{i}(t)
\end{equation} 
	as a positive constant given $\boldsymbol{U}(t)$, $\boldsymbol{Q}(t)$, and positive parameter $V$ at the beginning of time slot $t$.		
	}
\end{lemma}
\textbf{\textit{Proof}:} The proof is relegated to Appendix-A. 
	
\begin{algorithm}[!t]
 \caption{POTUS (Predictive Online Tuple Scheduling) in one time slot}
 \begin{algorithmic}[1] \label{algo}
	\STATE At the beginning of time slot $t$, the stream manager of each container $k$ collects instant system dynamics: $\boldsymbol{Q}(t)$ from its metric manager, and acquires ${U}_{k, *}(t)$. 
	\STATE \textbf{For} each instance $i \in \mathcal{I}_{K}(k)$ 
		\STATE $\ \ $
		Initialize the number of tuples to be sent in time slot 
		\\ $\ \ $
		$t$ as $\tilde{\gamma}_{i}(t) = 0$. 
		\STATE $\ \ $ 
		Pick out candidate instances $\mathcal{I}_{\text{cand}}(i)$ such that
				$$
					\mathcal{I}_{\text{cand}}(i) 
					\leftarrow 
					\left\{ 
						i' \in \mathcal{I}_{C}(c'), \forall\, c' \in n(i)
						\vert 
						l_{i, i'}(t) < 0
					\right\}.
				$$
		\STATE $\ \ $ 
		\textbf{If} $\mathcal{I}_{(\text{cand})}(i) = \emptyset$ then
			\STATE $\ \ \ \ $
			Distribute tuples in $Q^{\text{(rem)}}_{i, c'}(t, 0)$ among instances \\ 
			$\ \ \ \ $ in $\mathcal{I}_{C}(c')$ for $c' \in n(i)$.
		\STATE $\ \ $  
		\textbf{For} each subsequent component $c' \in n(i)$
			\STATE $\ \ \ \ $ 
				Initialize $\tilde{Q}_{i, c'}^{\text{(out)}}(t) \leftarrow {Q}_{i, c'}^{\text{(out)}}(t)$.\\
		$\ \ $ \%\% Making decisions for tuple scheduling
		\STATE $\ \ $ 
		\textbf{While} 
				$\tilde{\gamma}_{i}(t) < \gamma_{i}$ 
				and 
				$\mathcal{I}_{\text{cand}}(i) \neq \emptyset$:
				\STATE $\ \ \ \ $ 
					Pick such instance $i^{*}$ of component $c^{*}$ that 
					\\ $\ \ \ \ $
					$$
						i^{*} \in 
						\underset{i' \in \mathcal{I}_{\text{cand}}(i)}{\arg\min} \ \ 
						l_{i, i'}(t).
					$$
					\STATE $\ \ \ \ $ 
					Set $X_{i, i^*}(t) 
							\leftarrow 
							\min\{ {\gamma}_{i} - 
								   \tilde{\gamma}_{i}(t), 
								   \tilde{Q}_{i, c^{*}}^{\text{(out)}}(t)\}$.
					\STATE $\ \ \ \ $
					Update 
							$\tilde{\gamma}_{i}(t) 
							\leftarrow \tilde{\gamma}_{i}(t) + 
							X_{i, i^{*}}(t)$.
					\STATE $\ \ \ \ $
					Update 
							$\tilde{Q}_{i, c^{*}}^{\text{(out)}}(t) 
							\leftarrow 
							\max\{
								\tilde{Q}_{i, c^{*}}^{\text{(out)}}(t) - 
								X_{i, i^{*}}(t), 
								0
							\}$.
					\STATE $\ \ \ \ $
					Update $\mathcal{I}_{\text{cand}}(i)
							\leftarrow \mathcal{I}_{\text{cand}}(i)\backslash\{ i^{*} \}$.
	\STATE Update instances' queue backlogs according to (\ref{qupdate-oqueue-on-spout}) -- (\ref{qupdate-oqueue-on-bolt}).
\end{algorithmic}
\end{algorithm}

	In fact, problem (\ref{subpr-for-tuple-routing}) can be further decomposed on a per-instance basis and each instance makes its tuple scheduling decision independently.
	Nonetheless, it is not practical for implementation, because such per-instance optimization requires extra resource from each instance to keep its related states and undertake decision making, inducing considerable overheads. 
	Instead, it is more practical to implement the decision making for tuple scheduling on the stream manager in each container, since: 
	\begin{enumerate}[(i)]
		\item Stream manager itself manages the control of tuple transmission between instances within and across containers; hence it naturally keeps the instant information required to calculate (\ref{l-def}) such as $\boldsymbol{U}(t)$.
		\item The queue backlog sizes $\boldsymbol{Q}(t)$ of all its instances can be attained by direct interaction with the module that holds such information in each container, \textit{i.e.}, the metric manager in Apache Heron \cite{kulkarni2015twitter}.
		\item The number of instances in every container is often not very large, about ten on average \cite{numinstances}. 
	\end{enumerate}
	The above discussion suggests that it is a proper choice to conduct tuple scheduling by stream managers, 
	which also minimizes the impact of decision making process on instances' execution.
	In the following, we propose POTUS, an efficient and distributed scheme that solves problem (\ref{subpr-for-tuple-routing}) optimally and present its pseudocode in Algorithm \ref{algo}.

\textbf{Remark 1:}
	In particular, POTUS conducts the scheduling decisions in a distributed manner across containers. 
	By collecting all necessary information, the stream manager in each container can make scheduling decisions independently for its instances.
	The calculation of (\ref{l-def}) only requires the communication cost to other containers, its output queue backlog size, and the input queue backlog size on the target instance. 
	Particularly, for instances $i$ and $i'$, $l_{i, i'}(t)$ actually reflects the unit price of transferring a tuple from instance $i$ to instance $i'$. 
	{In $l_{i, i'}(t)$, the value of parameter $V$ can be viewed as the metric of relative importance of optimizing communication costs compared to stabilizing all queue backlogs in the system. 
	The larger the value of $V$, the more focus POTUS would put on minimizing communication costs of tuple scheduling. Equivalently, for each processing instance, POTUS is more prone to forwarding tuples to its successive instances that reside in the same or the nearest server, unless those instances are already heavily loaded. 
	Meanwhile, such scheduling may also lead to workload imbalance among instances and impede queue stability. 
	In contrast, the smaller the value of $V$, the more focus POTUS would put on forwarding tuples to maintain queue stability of instances. 
	However, to this end, some tuples should be sent across servers, leading to more communication costs. In brief, different values of $V$ can lead to different degrees of tradeoff between communication cost optimization and queue stability. 
	In practice, the choice of parameter $V$'s value depends on the operation objective of real systems. 
	A benchmark rule to decide the value of parameter $V$ is to set it as the ratio of the magnitude of total communication costs to the total queue backlog size. Under such a choice of $V$, POTUS would put equal focus on minimizing communication costs and queue stability.}	

{\textbf{Remark 2:}
POTUS incurs only little overhead for the acquisition of neighboring instances' dynamic information during each time slot. The detailed explanation is given as follows. 
Specifically, under POTUS, the scheduling of tuples during each time slot is conducted by stream managers in a distributed fashion. In particular, each stream manager aims to decide the forwarding of tuples that have been processed by instances within its container. To this end, during each time slot, the decision making of each stream manager requires the information about its instantaneous communication costs to stream managers in the other containers and queue backlog sizes of all instances in those containers. In practice, such information is collected and maintained by particular metric monitor processes (\textit{e.g.}, metric managers in Apache Heron). Meanwhile, such information is shared periodically among metric monitors and the update period is typically tens of milliseconds. Upon decision making, each stream manager can directly fetch such information from its corresponding metric monitor. Therefore, in practice, by setting the time slot length at the same magnitude of the update period, only little overhead would be incurred for the acquisition of information for decision making.	}
	
{\textbf{Remark 3:}
	In practice, given concrete information of application topology, it is viable to further improve the design of POTUS to achieve an even better performance. 
	Although the resulting design may impose extra constraints on tuple scheduling (\textit{e.g.}, tuple processing is correlated among successive instances), queue stability is still guaranteed. 
	This is because under POTUS, tuple scheduling is conducted across time slots in an adaptive manner. In particular, when data correlation leads to temporal workload increase in some processing queues, then in the next few time slots, tuples would be scheduled to other lightly loaded instances. 
	In this way, queue stability would be ensured in the long run.}

\subsection{Performance Analysis for POTUS}\label{Subsection: performance analysis}
According to Algorithm \ref{algo}, we analyze the time complexity of POTUS in one time slot. 
First, for each instance $i$, 
POTUS requires $O(C_{\text{max}} \cdot I^{C}_{\text{max}})$ time to form a candidate set from all its succeeding instances (line $4$), where $C_{\text{max}}$ denotes the maximum number of components in any application and $I^{C}_{\text{max}}$ denotes the maximum number of instances (parallelism) of any component. 
If the candidate set is not empty, then the tuple scheduling loop (line $12$-$18$) takes at most $C_{\text{max}} \cdot I^{C}_{\text{max}}$ iterations in the worst case, corresponding to the case with all successors of instance $i$ being scheduled in the process.
During each iteration, picking the target instance (line $13$) also requires at most $O(C_{\text{max}} \cdot I^{C}_{\text{max}})$ iterations to finish. 
Therefore, the time complexity for tuple scheduling loop is $O\left(
\left( C_{\text{max}} \cdot I^{C}_{\text{max}} \right)^{2} \right)$.
Consequently, the computational complexity of POTUS for each instance $i$ is $O((C_{\text{max}} \cdot I^{C}_{\text{max}})^{2})$.
To analyze the overall complexity, recall that each stream manager makes scheduling decisions for its residing instances, independent of the other stream managers.
Hence, with necessary information attained, 
stream managers can undertake the scheduling in parallel, 
with an overall complexity of $O(I^{K}_{\text{max}} \cdot (C_{\text{max}} \cdot I^{C}_{\text{max}})^{2})$, where $I^{K}_{\text{max}}$ denotes the maximum number of instances in any container. 
In practice, the number of components and the number of instances per container are usually no very large \cite{storm}\cite{numinstances}. 
Thus, POTUS trades off only little overheads for scheduling tuples with fine-grained control.

On the other hand, when all lookahead window sizes are zero ($W_i = 0$ for all $i$), 
\textit{i.e.}, no future information is available, POTUS degenerates to the classical Lyapunov optimization algorithm and achieves an $\left[O(V), O(1/V)\right]$ trade-off between the time-average total queue backlog size and the time-average total communication cost via a tunable parameter $V$, while guaranteeing the stability of queue backlogs in the system. 
In particular, given the value of $\beta$, let $\bar{\Theta}^*$ denote the optimal value of problem (\ref{problem-def-small-time-scale}), the we have the following theorem.
\begin{theorem}
	\textit{
	Suppose that $h(0) < \infty$ and, given the processing capacities of instances and their placement, there exists an online scheme which ensures that, for each instance, the mean arrival rate is smaller than its mean service rate. Then, under POTUS without prediction ($W_i=0$ for each spout instance $i$), there exist constants $B > 0$ and $\epsilon > 0$ such that
	\begin{equation}
	    \begin{split}
	    \limsup_{T\rightarrow \infty} \frac{1}{T}&\sum_{t=0}^{T-1} \mathbb{E} \left\{ \left( \Theta(t) \right) \right\} \leq \bar{\Theta}^* + \frac{B}{V} \\ 
	    \end{split},
	\end{equation}
	\begin{equation}
		\frac{1}{T}\sum_{t=0}^{T-1} \mathbb{E} \left\{ h(t) \right\} \leq \frac{V \bar{\Theta}^{*}}{\epsilon} + \frac{B}{\epsilon}.
	\end{equation}
	}
\end{theorem}
\textbf{\textit{Proof:}} The proof is relegated to Appendix-B. 

The above theorem implies that, 
by solving problem (\ref{subpr-for-tuple-routing}) over time slots, POTUS can achieve near-optimal long-term time-average total communication costs with an $O(1/V)$ optimality gap for problem (\ref{problem-def-small-time-scale}). 
By increasing the value of $V$, POTUS can reduce the optimality gap but at the cost of an increased total queue size. 
According to Remark 1 in Section 4.1, a large value for $V$ will incentivize stream managers to distribute tuples among the instances of successive components with the minimum communication costs. 
However, instances with limited processing capacities in some containers may become hot spots with an ever-increasing queue size. 
In contrast, a smaller value of parameter $V$ conduces to more balanced queue sizes among instances but that may require more cross-container transmissions and hence greater communication costs.
Moreover, when future information is available, POTUS can achieve an even better performance with predictive scheduling, as shown in Section 5.

\section{Simulation}  \label{sec: simulation}
In this section, we evaluate the performance of POTUS under various settings and compare it with the built-in scheme in Apache Heron to investigate the fundamental benefits of predictive scheduling and the impact of mis-predictions. 
In the following subsections, we first illustrate the basic settings of our simulations, then discuss our simulation results in detail.

\subsection{Basic Settings}

We construct two stream processing systems based on two widely adopted topologies, Jellyfish\cite{jellyfish} and Fat-Tree\cite{fat-tree}.
Each of them contains $24$ switches and $16$ servers that host stream processing applications. 
  
{\textbf{Stream Processing Applications:}}
In the simulation, we deploy five data stream processing applications with commonly adopted topologies\cite{kulkarni2015twitter}\cite{t-storm}\cite{peng2015r}. 
Particularly, each application has a topology depth varying from $3$ to $5$ and a number of components ranging from $3$ to $6$.
Instances of the same component have the identical processing capacity ranging from $3$-$5$ tuples per time slot. 

{\textbf{Instance Placement:}}
For instance placement, the mapping from instances to containers is determined by T-Heron, a placement scheme which is adapted from T-Storm\cite{t-storm}.
Given a new application, T-Heron sorts all its instances by their descending order of (incoming and outgoing) tuple traffic rate. Then it iteratively assigns each instance to one of the available containers with minimum incremental traffic. 

{\textbf{Traffic Workloads:}}
We conduct trace-driven simulations with tuple arrival measurements drawn from real-world network systems\cite{benson2010network}.  
In addition, we also conduct simulations where tuple arrivals follow Poisson distribution, with the same arrival rate as in trace-driven cases.

{\textbf{Prediction Settings:}}
The traffic of different applications often varies in predictability. 
By fixing the average window size as $W$, the prediction window size for each application is set by sampling uniformly from $[0, 2\!\times\!W]$ at random.
We evaluate cases with perfect and imperfect prediction.
For \textit{perfect prediction}, future tuple arrivals in the time window are assumed perfectly predicted and can be pre-served. 
For \textit{imperfect prediction}, we implement five prediction schemes that predict tuples yet to arrive (all with $W\!=\!1$), with mean-square error (MSE) varying from $10.37$ to $22.54$, including:
1) \textit{Kalman filter}\cite{chui2017kalman}; 
2) distribution estimator (\textit{Distr}), which predicts the number of arriving tuples in each future time slot by independent sampling from the empirical distribution of the frequency of arriving tuple numbers  in the past time slots;
3) \textit{Prophet} \cite{taylor2018forecasting}, Facebook's time-series prediction procedure;
4) moving average (\textit{MA}) and 5) exponentially weighted moving average (\textit{EWMA})\cite{box2015time}.
Then we consider two basic types of mis-prediction.
One is \textit{true negative}, \textit{i.e.}, when an actual tuple is not predicted to arrive and hence not pre-served before its arrival. 
The other is \textit{false positive}, \textit{i.e.}, when a tuple that does not exist is predicted to arrive; in this case, processing such tuples consumes extra system resources.
To investigate the limits of predictive scheduling, we compare predictive scheduling with perfect prediction against two extremes of the spectrum: 
1) all actual tuple arrivals fail to be predicted; 
2) the actual arrivals are correctly predicted with different levels of false-positive prediction.

{\textbf{Compared Baselines for Tuple Scheduling:}
All instances are assumed stateless so that for any component, 
any of its instances can provide identical processing to the incoming tuples. 
We compare POTUS with Shuffle, \textit{i.e.}, the default tuple scheduling scheme in Heron, which dispatches each tuple to one of the instances of the corresponding next components uniformly at random.

{\textbf{Metric of Response Time:}}
For each tuple, its response time is counted as the number of time slots from 
its actual arrival to the last completion of its descendant tuples.
If a tuple is pre-served before its actual arrival, 
then it will be responded instantly upon its arrival and thereby incurring a negligible response time.

\subsection{Performance Evaluation and Analysis}

We fix the instance placement and evaluate the performance of POTUS under both perfect and imperfect predictions.

\subsubsection{Performance Evaluation under Perfect Prediction}

To investigate the benefits of predictive scheduling, 
we compare POTUS with zero ($W\!=\!0$) and non-zero lookahead window sizes under different settings. 
Note that the former is a special case of POTUS without future information. 

\begin{figure}
  \centering 
  \subfigure[Jellyfish Topology]{ 
	\includegraphics[scale=.15]{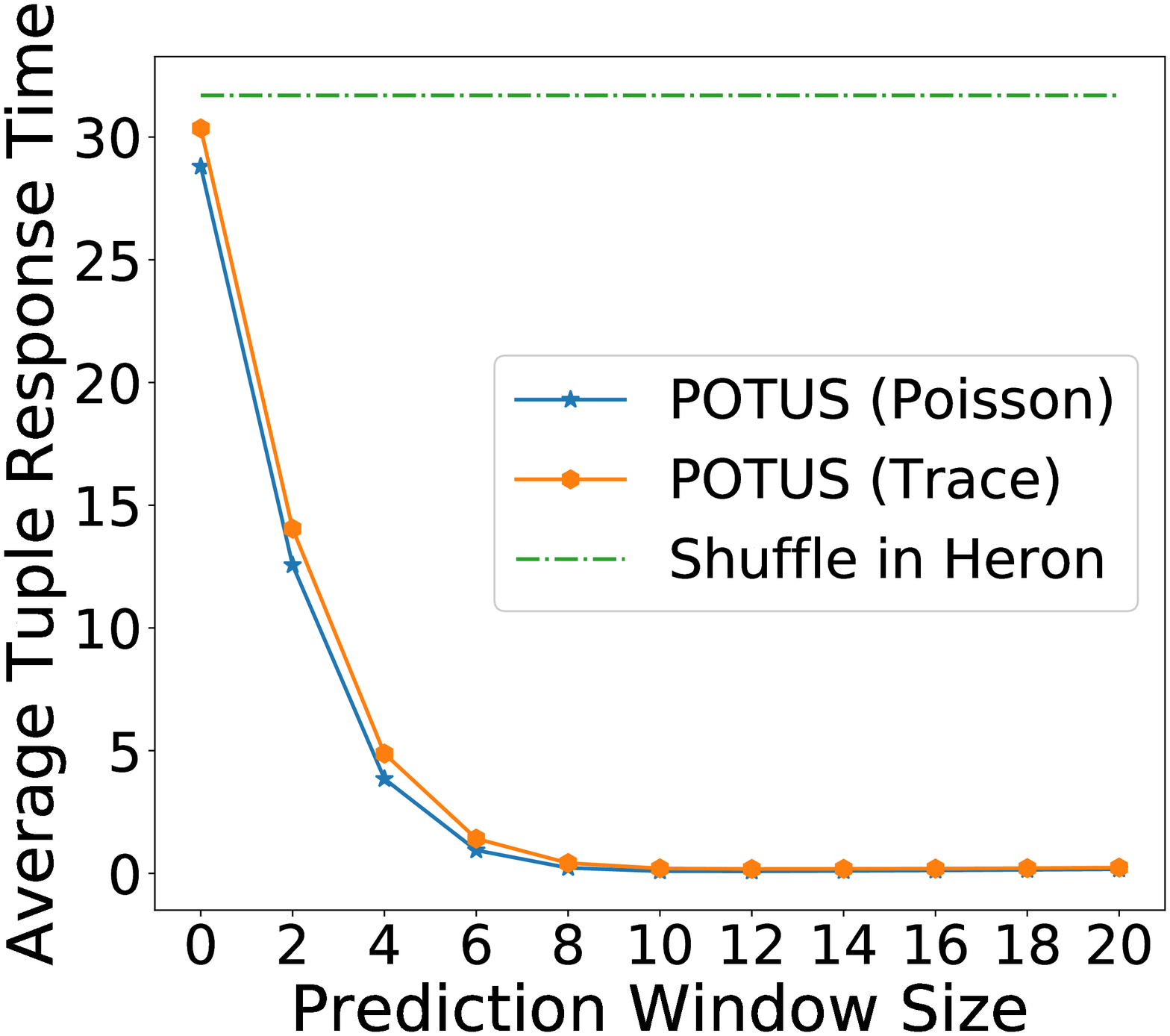}
  } 
  \hspace{0.4em}
  \subfigure[Fat-Tree Topology]{
  	\includegraphics[scale=.15]{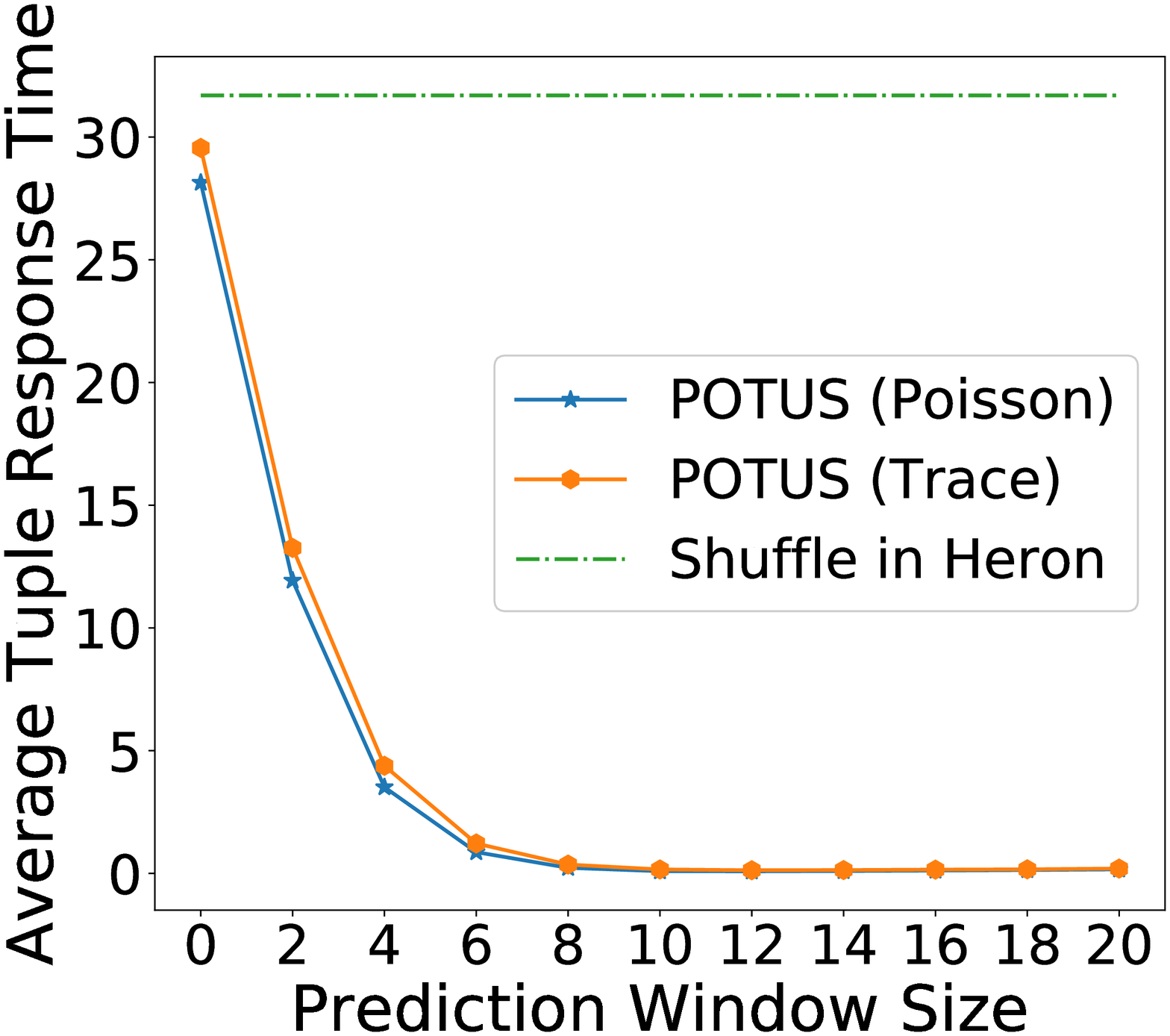}
  } 
\caption{Average response time vs. look ahead window size with Poisson and trace arrival processes under different topologies, given $V=3$.}
\label{win-vs-resp-perfect}
\end{figure}

\begin{figure*}[!t]
  \subfigure[Jellyfish Topology]{ 
	\includegraphics[scale=.18]{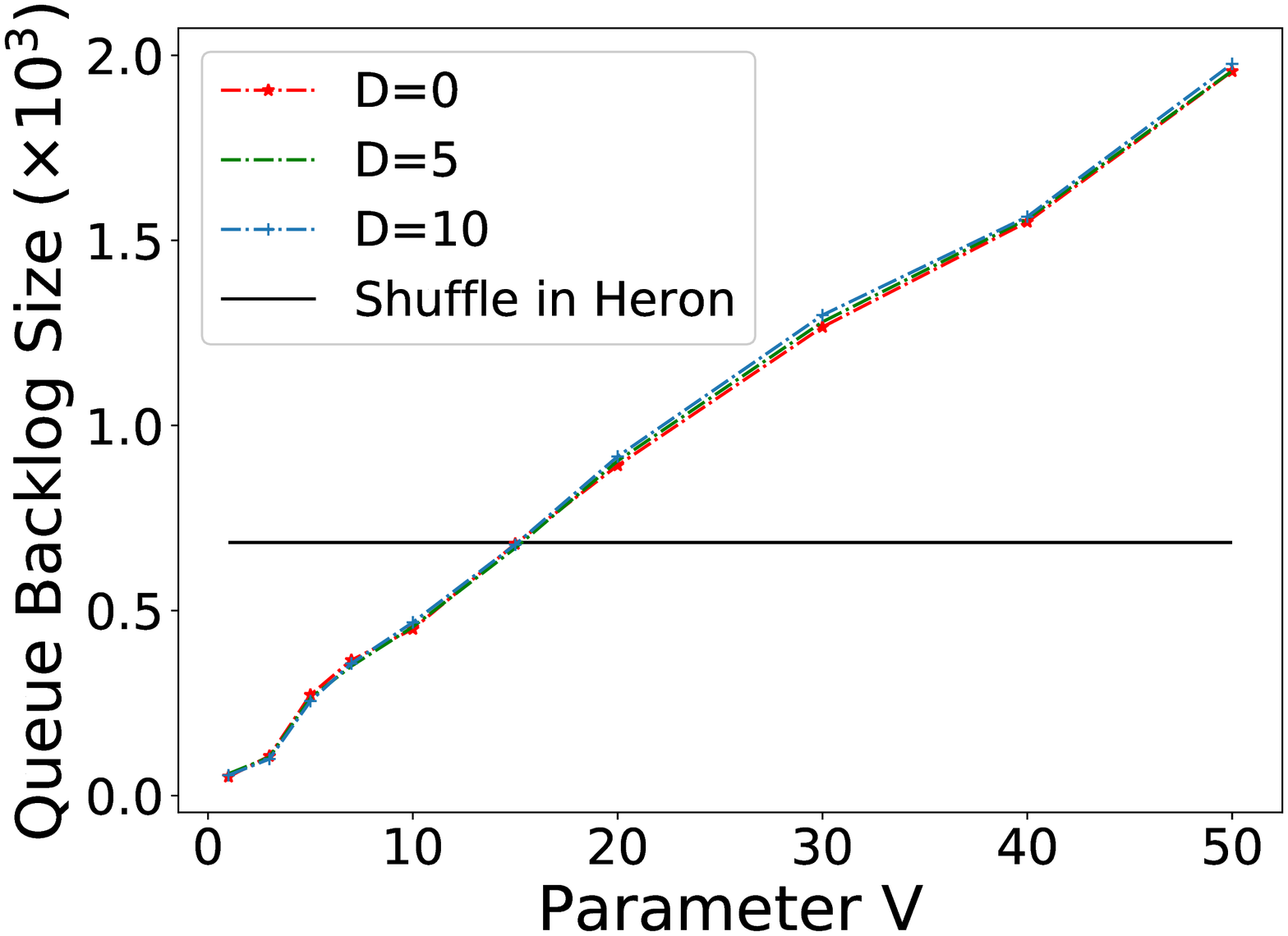}
  } 
  \subfigure[Fat-Tree Topology]{
  	\includegraphics[scale=.16]{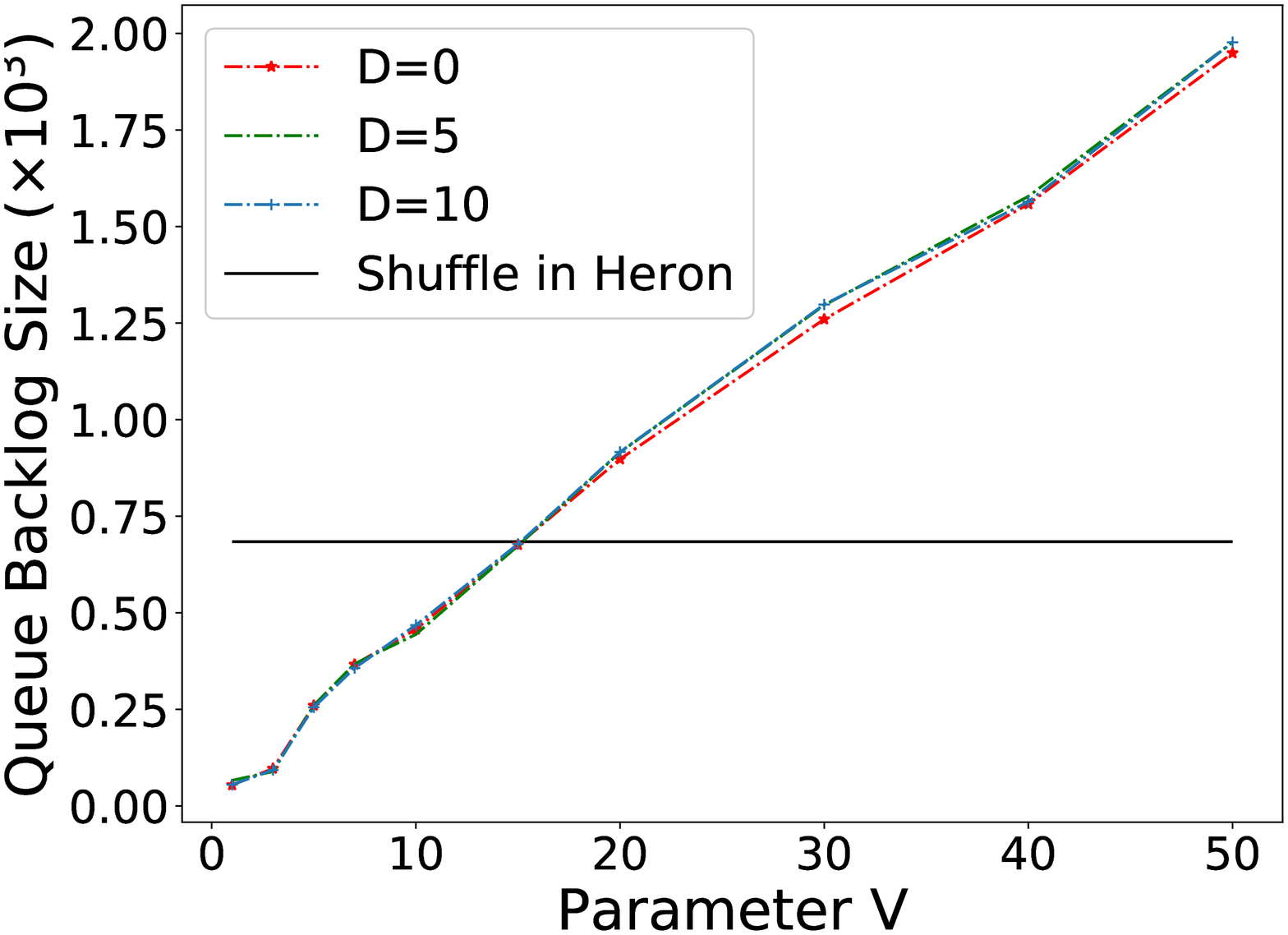}
  } 
  \centering 
  \subfigure[POTUS ($W=0$)]{ 
	\includegraphics[scale=.14]{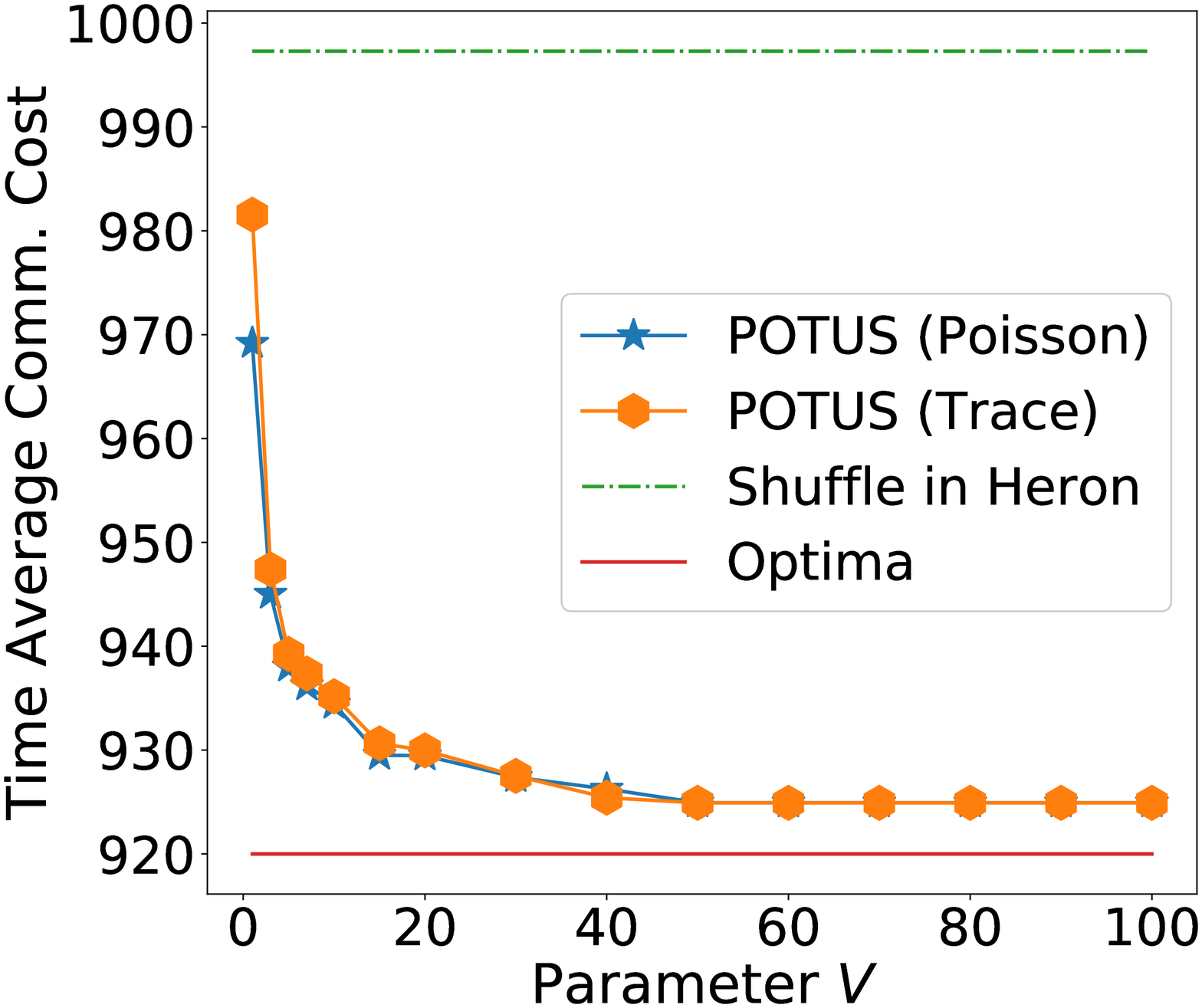}
  } 
  \hspace{0.5em}
  \subfigure[POTUS ($W=10$)]{
  	\includegraphics[scale=.14]{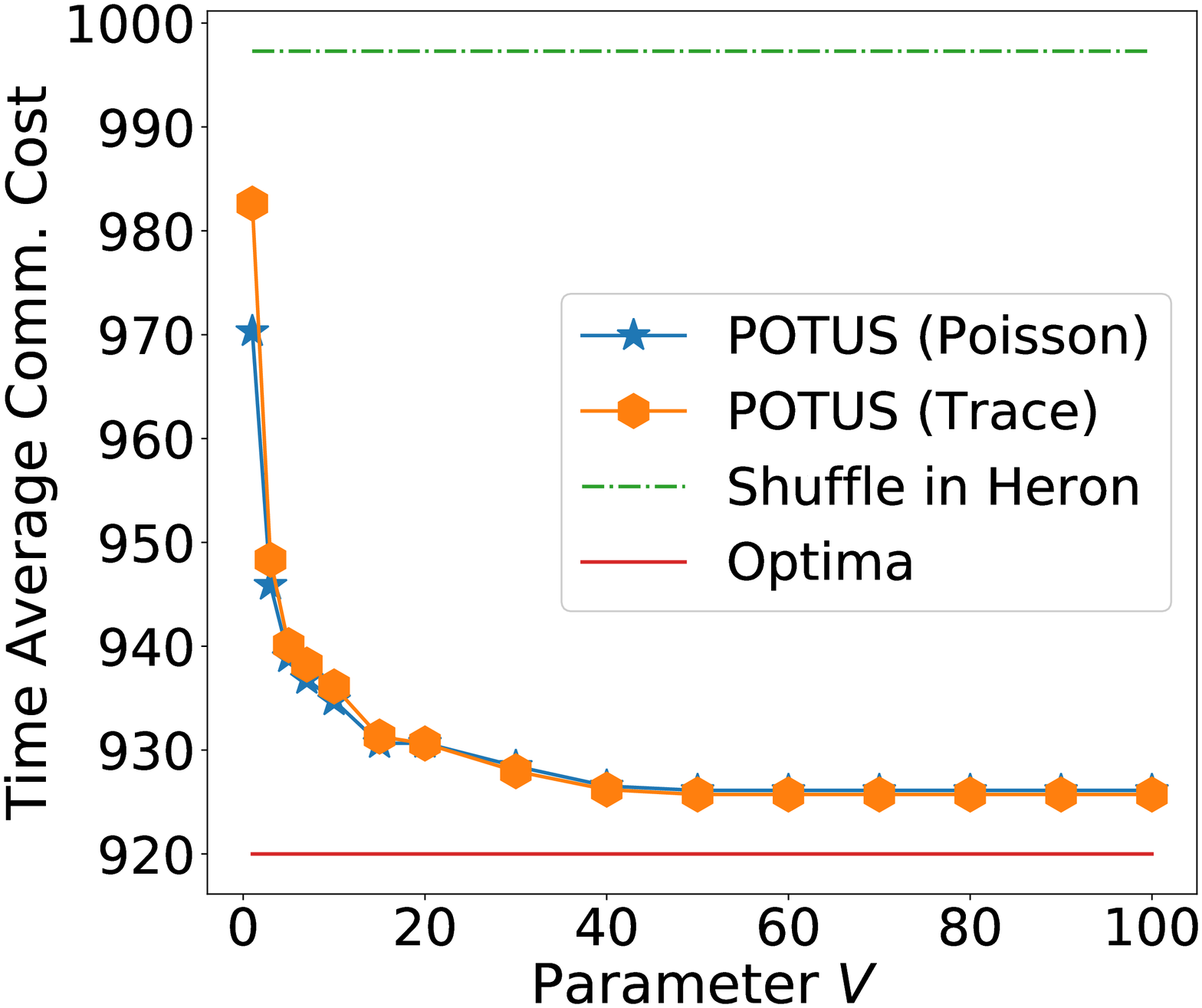}
  } 
  \caption{
Performance of POTUS under various settings with perfect prediction. 
}
\label{figures: various plots}
\end{figure*}

{\textbf{Average response time vs. lookahead window size $W$:}}
Figures \ref{win-vs-resp-perfect}(a) and \ref{win-vs-resp-perfect}(b) show the results under Jellyfish and Fat-Tree topology, respectively. 
We take the curve of Jellyfish under Trace arrival as an example. 
As the value of window size $W$ increases from $0$ to $6$, 
POTUS shortens the average response time from $31.4 ms$ to $1.5 ms$. 
As the value of $W$ continues increasing, 
the marginal reduction in average response time diminishes.
Eventually, the response time remains stably at $0.5 ms$. 
Likewise, curves under Fat-Tree topology also exhibit similar trends.
On the other hand, Shuffle scheme incurs an average response time about $5\%$ higher than POTUS without future information ($W=0$). 
This is reasonable since Shuffle implicitly assumes the homogeneity of instances' processing capacities 
while POTUS exploits instant system dynamics. 

\textit{Insight:}
Results from Figure \ref{win-vs-resp-perfect} highlight the benefits of predictive scheduling in shortening the average tuple response time. 
Moreover, mild-value of future information is sufficient to effectively reduce tuple response time to nearly zero. 
The marginal benefits then diminish with more future information due to the limit in system service capacity.

{\textbf{Time-average total queue backlog size under different settings:}}
Figures \ref{figures: various plots}(a) and \ref{figures: various plots}(b) show how the total system queue backlog size changes under different topologies, window sizes, and the values of parameter $V$. 
From both plots, it turns out that: 
1) the Shuffle scheme incurs a constant backlog size since its has nothing to do with the value of $V$; 
2) with different window sizes, 
POTUS incurs a linearly ascending trend in the queue backlog size as the value of $V$ grows from $1$ to $50$. 
Particularly, under Fat-Tree topology, as the value of $V$ becomes greater than $16$, the system cumulates more queue backlog by POTUS than the Shuffle scheme in Heron. 
This is because a larger value of parameter $V$ leads instances to forward more tuples to those with low communication costs among successive instances,
resulting in more unbalanced workloads across instances. 
In contrast, with a small value of $V$, POTUS focuses on steering tuples between instances to stabilize queue backlogs and incurs more balanced workloads. 
This implies that, in practice, one should greedily choose a small value of parameter $V$. However, as we shall see later, the choice of $V$ also affects how communication costs change and thus a trade-off must be made.

Considering similarities of the results under different topologies and arrivals processes, in the following, 
we focus on the results under Fat-Tree topology with trace-driven arrivals.

{\textbf{Time-average communication costs vs. $V$:}}
Figures \ref{figures: various plots}(c) and \ref{figures: various plots}(d) evaluate the performance of POTUS in terms of time-average total communication cost as the value of parameter $V$ varies from $1$ to $100$,
with $W=0$ and $W=10$, respectively.

Figure \ref{figures: various plots}(c) shows that POTUS outperforms the Shuffle scheme in Heron by up to $7.6\%$ reduction in the communication cost. 
Besides, for the curves of POTUS, we also see the dramatic descending trend of communication costs by increasing the value of $V$ from $1$ to $100$.
Particularly, when the values of $V$ vary from $1$ to $40$, the time-average communication cost decreases by $5.8\%$; 
however, if further increasing $V$, we see the cost reduction stops and remains constant after $V=50$, leaving a $0.5\%$ gap to the optimal cost. 
Such an optimality gap is the price to take for stabilizing the system queue backlogs. 
To maintain queue stability, 
they would not choose their proximal instances that are heavily loaded.
Instead, they turn to other less loaded instances with larger communication costs, 
thereby incurring sub-optimal total costs. 
The descending trend remains the same for both $W=0$ and $5$. 
This shows that under perfect prediction, predictive scheduling incurs almost no extra system costs. 

\textit{Insight:}
Figures \ref{figures: various plots}(a) - \ref{figures: various plots}(d) show that, by increasing the value of parameter $V$, the total queue backlog size would increase and the total communication costs would decrease. 
Therefore, parameter $V$ actually determines the trade-off between communication costs and queue backlog size in the system, as shown by our previous analysis in Section \ref{Subsection: performance analysis}.
By choosing a relatively small value for parameter $V$, \textit{e.g.}, $V \in [5, 15]$ in our simulation, 
POTUS can attain both lower communication costs and smaller queue backlog size.
In practice, the value of $V$ can be tuned according to particular objectives in system design.

\begin{figure*}[!h]
  \centering 
  \subfigure[Communication Cost with Parameter $V$]{ 
	\includegraphics[scale=.2]{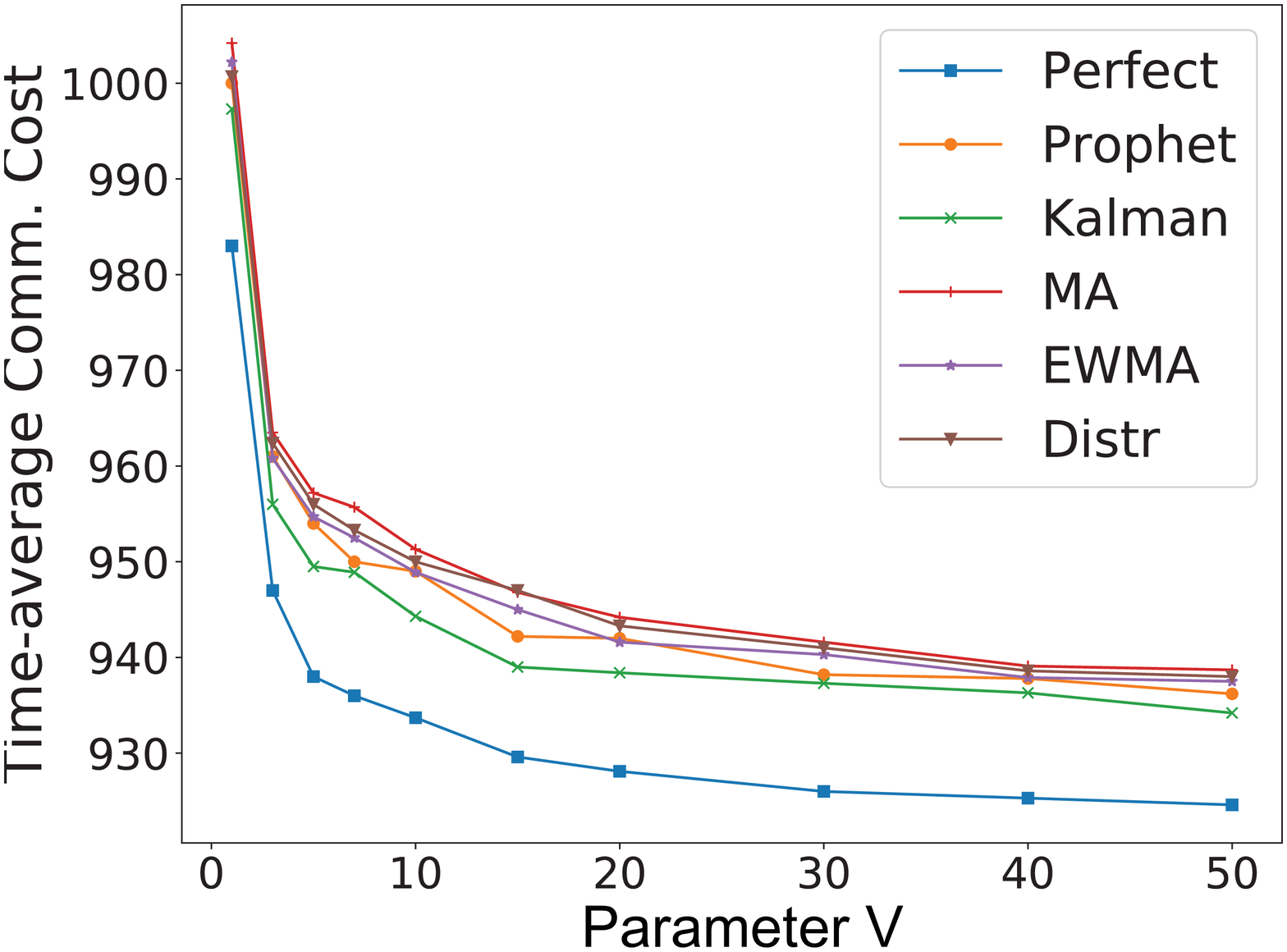}
  } 
  \hspace{1em}
  \subfigure[Response Time with Parameter $V$]{
  	\includegraphics[scale=.2]{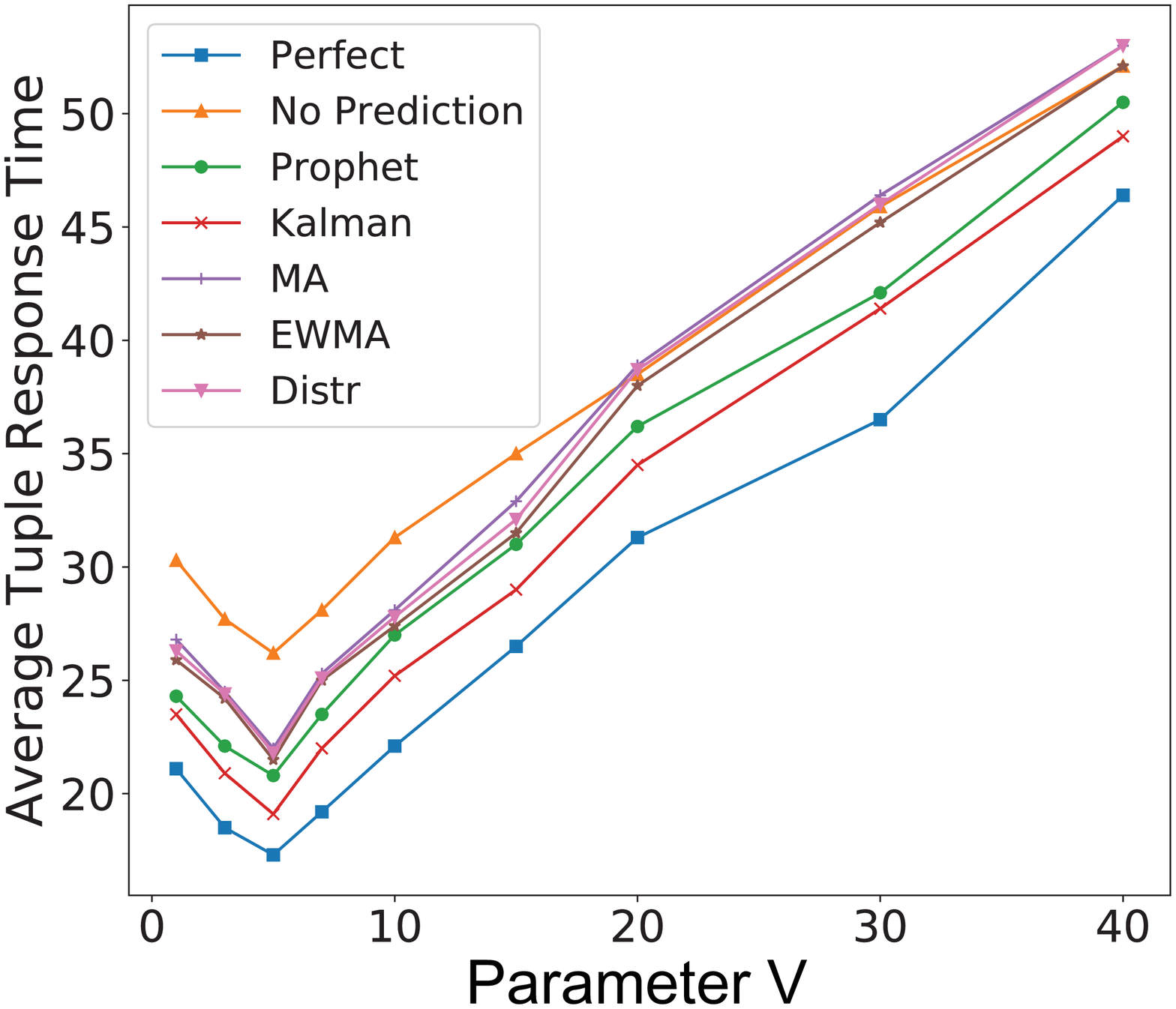}
  }
  \hspace{1em}
  \subfigure[Response Time with Window Size]{ 
	\includegraphics[scale=.28]{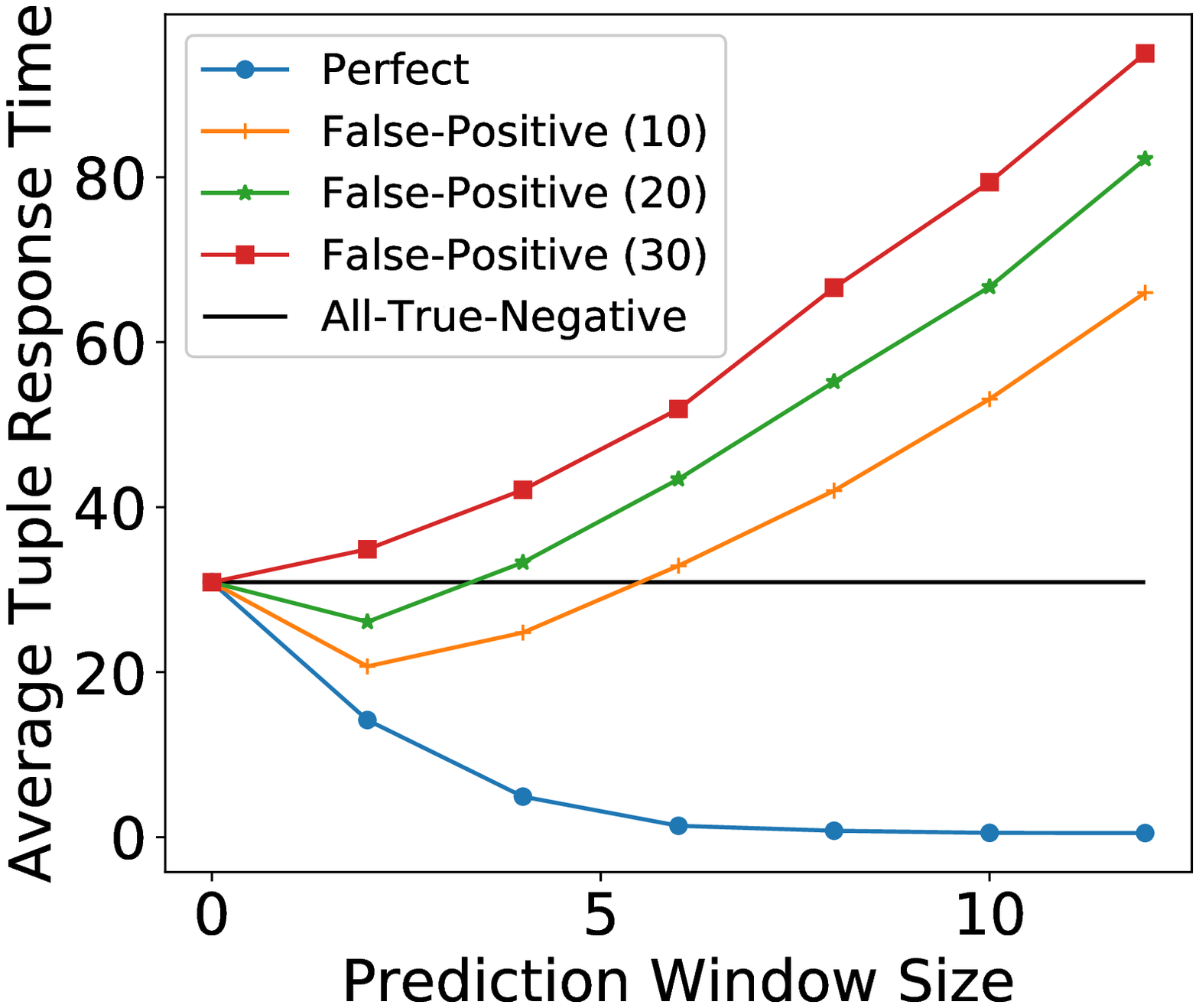}
  }
\caption{
	Average tuple response time under perfect prediction and different cases of mis-prediction. 
}
\label{mis-prediction}
\end{figure*}

\subsubsection{Performance Evaluation under Imperfect Prediction}
{In practice, prediction errors are inevitable. Due to such inaccuracies, two cases would occur. On one hand, the system may wrongly allocate extra resources to pre-serve those falsely alarmed tuples (\textit{a.k.a.} false-positive). As a result, the response time of data streaming services would be further extended. On the other hand, failing to predict part of tuple arrivals (\textit{a.k.a.} true-negative) would lead to resource under-provisioning. Accordingly, such tuples do not enjoy pre-service and hence no delay reduction. Consequently, the coexistence of such two cases would offset the benefit of predictive scheduling. The performance would even worsen as the degree of inaccurate prediction increases.}
In this subsection, we first evaluate the performance of POTUS with five imperfect prediction schemes, including \textit{Kalman}, \textit{Distr}, \textit{Prophet}, \textit{MA}, and \textit{EWMA}, which may incur mis-prediction such as false-positive and true-negative predictions of tuple arrivals. 
Meanwhile, we consider the performance of POTUS under perfect prediction as the benchmark to investigate such prediction schemes. 
Since some of the schemes (\textit{e.g.}, MA and EWMA) are only able to make short-term prediction for the next time slot, for the fairness of comparison, we set the value of window size $W$ as $1$ in all cases. 
\footnote{Moreover, given temporal variations of data stream dynamics, short-term prediction (\textit{e.g.}, lookahead prediction only for the next time slot) is often the primary choice in practice.}
Besides, we also investigate the fundamental benefits of predictive scheduling in face of mis-prediction. 

{\textbf{Time-average communication costs vs. $V$:}}
Figure \ref{mis-prediction}(a) shows the time-average communication cost induced by POTUS with perfect prediction and five prediction schemes. 
We see a rapid reduction of communication costs under all schemes. 
With perfect prediction, the communication costs are even lower than the other five prediction schemes. 
The superfluous cost is caused by mis-prediction, especially false-alarm, which misleads the system to allocate extra resources to process tuples that do not exist.
Therefore, improving prediction accuracy conduces to lower communication costs.  

{\textbf{Average response time vs. $V$:}} 
Figure \ref{mis-prediction}(b) shows how average tuple response time changes under different choices of parameter $V$. We make the following observations. 

First, as the value of $V$ increases from $0$ to $5$, the average response time decreases by about $20\%$ and then a linear growth of response time thereafter. 
Such a trend change in response time is mainly determined by two factors: tuples' queueing delay and the communication cost. 
We know that increasing the value of parameter $V$ leads to the reduction in communication costs and growth in queue backlog size. 
Recall that by \textit{Little's theorem}\cite{little1961proof}, a greater queue backlog size implies a longer queueing delay for tuples. 
From Figure \ref{mis-prediction}(a), we see a significant reduction in the communication cost as the value of parameter $V$ rises from $1$ to $10$. 
Therefore, in this phase, the reduction in communication costs dominates and results in decreased response time. 
However, as the value of $V$ continues to increase, the reduction in communication cost diminishes while queue backlogs keep accumulating. Consequently, the increase in queueing delay dominates and induces the growth in response time.  

Second, compared to the perfect prediction case, those five schemes generally incur a longer response time.
Such extra response time is due to mis-prediction with both false positive and true negative.
In particular, false-alarmed tuples may be pre-served by the systems and take up the queue backlogs, incurring longer queueing delays for tuples that really need processing; meanwhile, miss-detected tuples take no benefits from pre-serving and hence their response time will not be shortened accordingly. 
In the meantime, compared to POTUS without prediction (\textit{No Prediction}), the five prediction schemes, although with mis-prediction, still benefit from pre-serving tuples by exploiting limited future information, 
by an up to $16.5\%$ reduction in response time. However, such benefits vanish as the value of parameter $V$ increases and exceeds $20$. 
The reason is that, for these prediction schemes, 
larger values of $V$ incur greater queue backlog sizes and more mis-predicted tuples. The accumulation of such tuples further extends the queueing delay and hence a longer response time than the non-prediction case. 

\textit{Insight:} 
The above results reveal two sides of predictive scheduling. 
On the upside, exploiting future information does help to shorten tuple response time. 
The downside is that, because of the unavoidable mis-prediction, the benefit of predictive scheduling will be gradually offset by the loss in terms of longer queueing delays, which are mainly caused by accumulated mis-predicted tuples with excessively leveraged future information, \textit{e.g.}, with a larger lookahead window. 
In practice, to enjoy the benefits brought by predictive scheduling, the balance must be carefully treated. 

{\textbf{Average response time vs. lookahead window size $W$:}}
In Figure \ref{mis-prediction}(c), we further consider two extreme cases of mis-prediction. 
One is when all actual tuple arrivals fail to be predicted, denoted by \textit{All-True-Negative}.
In fact, \textit{All-True-Negative} is equivalent to the case without prediction since no tuples are predicted to arrive. 
The other is when the actual arrivals are correctly predicted, but also with some extra arrivals falsely alarmed.
We denote such a case by \textit{False-Positive ($x$)}, where $x$ is the average number of false-positive tuples. 
Note that any case of mis-prediction is equivalent to some superposition of these two extremes. 
By fixing $V=1$ and varying the value of $x$ from $10$ to $30$, we acquire the following results. 

First, we see that, with perfect prediction, POTUS effectively reduces tuple response time down to near zero, as the prediction window size grows from $0$ to $10$.
\textit{All-True-Negative}, as anticipated, remains constant in response time.
Meanwhile, \textit{False-Positive} exhibits various performances with different values of $x$.
In particular, when $x=10$, the response time declines as the window size grows from $0$ to $2$, and goes up but remains lower than \textit{All-True-Negative} before the window size reaching $6$. 
From Figure \ref{mis-prediction}(b), we infer that mis-prediction, especially false-positive requests, contribute to the increase in queueing delay and thus prolongs response time. 
Such reduction in response time actually attributes to the advantage of predictive scheduling, 
\textit{i.e.}, by exploiting surplus system resources in present time slot to pre-serve future requests and achieving more balanced workloads in temporal dimension. 
Nonetheless, as the window size continues to increase from $6$ to $10$, more and more false-positive requests will be admitted and pre-served by the system. 
Finally, the response time is dominated by the ever-increasing queueing delay, while the benefit of predictive scheduling vanishes. 
By increasing the number of false-positive tuples from $10$ to $30$, predictive scheduling no longer conduces to shortening response time. 
Instead, with excessively pre-served false-positive tuples, it results in an even longer response time.

{
Our results show that in general, the two types of mis-prediction (\textit{i.e.}, \textit{All-True-Negative} and \textit{False-Positive}) would lead to a longer response time with only mild changes in communication costs.
Specifically, in the former case, those affected tuples would not benefit from predictive scheduling. Intuitively, the transmission and processing of such tuples are just deferred in the temporal dimension. Therefore, their mis-prediction would lead to a longer response time than that under perfect prediction with only negligible changes in the time-averaged communication costs. In the latter case, the accumulation and processing of mis-predicted tuples may consume extra system resources. As a result, the tuple response time would be extended with a mild increase in the time-averaged communication costs. 
This is because the number of treated tuples to transmit for each instance during each time slot is mainly determined and limited by its processing capacity. Hence, the change in time-averaged communication costs is also limited. This also implies that with mis-predicted tuples, the transmission of some actual tuples would be deferred, thereby leading to a longer response time. 
To mitigate such issues, in practice, POTUS can be adapted by giving priority to jobs already in the system to further improve application performances.}

\textit{Insight:} 
On one hand, even with mis-prediction, short-term prediction is still helpful to shorten tuple response time. Note that such results also verify the robustness of POTUS against prediction errors. 
On the other hand, to exploit the power of predictive scheduling, one should adopt the prediction scheme with a small variance in the estimate of tuple arrivals, 
so that false-positive arrivals would not offset the benefit of reduced response time.

\section{Conclusion}  \label{sec: conclusion}
In this paper, we studied the problem of tuple scheduling in Apache Heron, with a systematic investigation of the fundamental benefits of predictive scheduling in such systems.
Through a careful system modeling and problem formulation, we proposed \textit{POTUS}, a distributed scheme that schedules tuples with pre-service in an online fashion.
Our theoretical analysis and simulation results verified the effectiveness of POTUS in achieving near-optimal communication costs and ultra-low tuple response time. 
Moreover, our results showed that only mild-value of future information suffices to lead to a notable reduction in tuple response time, even in the presence of mis-prediction.



\appendices
\section{Proof of Lemma 1}
We define the quadratic Lyapunov function as
\begin{equation}\label{def-Lfunc}
    \begin{split}
	\displaystyle & 
	L(\boldsymbol{Q}(t)) \triangleq \\
	& \frac{1}{2}
	\bigg\{
		\sum_{i \in \mathcal{I}} 
		\left[
			Q^{(\text{in})}_{i}(t)	
		\right]^{2}
		+ \beta 
		\sum_{i \in \mathcal{I}}
		\sum_{c' \in n(i)}
		\left[
			Q_{i, c'}^{\text{(out)}}(t)
		\right]^{2}
	\bigg\}
	\end{split}
\end{equation}
and the Lyapunov drift for two consecutive time slots as
\begin{equation}
	\begin{split}
		\displaystyle \Delta (\boldsymbol{Q}(t))  \triangleq &~ \mathbb{E} \left\{ L(\boldsymbol{Q}(t+1)) - L(\boldsymbol{Q}(t)) \bigg\vert \boldsymbol{Q}(t)\right\}, \\
	\end{split}
\end{equation}
which measures the conditional expected successive change in queues' congestion state.
To avoid overloading any queue backlogs in the system, it is desirable to make the difference as low as possible. 
However, striving for small queue backlogs may incur considerable communication cost and computation cost. 
Hence, we should jointly consider both queueing stability and the consequent system costs.
Given decision $\boldsymbol{X}(t)$, 
we define the drift-plus-penalty function as
\begin{equation}
	\begin{split}\label{drift-plus-penalty}
		\displaystyle \Delta_V (\boldsymbol{Q}(t))  \triangleq &~ \mathbb{E} \left\{ L(\boldsymbol{Q}(t+1)) - L(\boldsymbol{Q}(t)) \bigg\vert \boldsymbol{Q}(t)\right\}\\
		& + V \mathbb{E} \left\{ \Theta\left( \boldsymbol{X}(t) \right) \big \vert \boldsymbol{Q}(t) \right\},
	\end{split}
\end{equation}
where parameter $V$ is a positive constant that quantifies the balance between stability and system cost reduction. 

Next, we show how problem (\ref{problem-def-small-time-scale}) is transformed into (\ref{problem_per_time_slot}) in detail. 
	According to (\ref{drift-plus-penalty}), the drift-plus-penalty term is 
	\begin{equation}
	\arraycolsep=1.1pt\def\arraystretch{1.7}
		\begin{array}{cl}
		    & \Delta_V (\boldsymbol{Q}(t)) \\
		    = & ~ \mathbb{E} \left\{ L(\boldsymbol{Q}(t+1)) - L(\boldsymbol{Q}(t)) + V \cdot \Theta(\boldsymbol{X}(t)) \bigg \vert \boldsymbol{Q}(t) \right\}.
	    \end{array}
	\end{equation}
Recall the definition in (\ref{def-Lfunc}), thus we can write $\Delta_V (\boldsymbol{Q}(t))$ as
	\begin{equation}\label{drift_plus_penalty_expand1}
	\arraycolsep=1.1pt\def\arraystretch{1.5}
		\begin{array}{cl}
			& \displaystyle \Delta_V (\boldsymbol{Q}(t)) \\
			= & \displaystyle \frac{1}{2} 
			\mathbb{E}\Big\{ 
				\sum_{i \in \mathcal{I}}
				\left[
					\left( Q^{(\text{in})}_{i}(t+1) \right)^{2} -
					\left( Q^{(\text{in})}_{i}(t) \right)^{2}
				\right]
				\bigg\vert \boldsymbol{Q}(t)
			\Big\} \\
			+ & \displaystyle 
			\frac{\alpha}{2} 
			\mathbb{E}\big\{ 
			\sum_{i \in \mathcal{I}}
			\sum_{c' \in n(i)}
			\left[
				\left( Q^{(\text{out})}_{i, c'}(t+1) \right)^{2} - 
				\left( Q^{(\text{out})}_{i, c'}(t) \right)^{2}
			\right] \\
			& \displaystyle
			\left.
			\bigg\vert \boldsymbol{Q}(t) \right\} + \mathbb{E} \left\{ V \Theta(\boldsymbol{X}(t)) \bigg \vert \boldsymbol{Q}(t) \right\}.
		\end{array}
	\end{equation}
Accordingly, we have 
%
	1) For spout instance $i$, regarding its output queue backlog to component $c' \in n(i)$, we have
	\begin{equation} 
	\arraycolsep=1.1pt\def\arraystretch{1.5}
	\begin{array}{rl}
		& \displaystyle 
		   \left[ Q^{\text{(out)}}_{i, c'}(t+1) \right]^2
			 \\
		=& \displaystyle 
			   \Big\{ 	
				   	\Big[
						Q^{\text{(out)}}_{i, c'}(t) - 
						\sum_{i' \in \mathcal{I}_{C}(c')} 
						X_{i, i'}(t)
					\Big]
					+ \lambda_{i, c'}(t + W_{i} + 1)
			   \Big\}^2 \\
		\leq & \displaystyle 
			   \left[ Q^{\text{(out)}}_{i, c'}(t) \right]^2 + 
			   \Big[
				   	\sum_{i' \in \mathcal{I}_{C}(c')} 
				   	X_{i, i'}(t)
			   \Big]^2 +
			   \Big[ \lambda_{i, c'}(t + W_{i} + 1) \Big]^2 
			   \nonumber
		\end{array}
	\end{equation}
	\begin{equation}\label{q_update_bound_on_spout_oqueue}
		\arraycolsep=1.1pt\def\arraystretch{1.5}
		\begin{array}{rl}	   
		& \displaystyle 
			   + 2 Q^{\text{(out)}}_{i, c'}(t)
			   \Big[ 
			   	   \lambda_{i, c'}(t + W_{i} + 1)
				   - \sum_{i' \in \mathcal{I}_{C}(c')} 
				   X_{i, i'}(t)
			   \Big].
	\end{array}
	\end{equation}
	
	2) For bolt instance $i$, regarding its input queue, we have
	\begin{equation} 
	\arraycolsep=1.1pt\def\arraystretch{1.5}
	\begin{array}{cl}
	 	& \displaystyle 
	 	\left[ Q^{\text{(in)}}_{i}(t+1) \right]^2 \\
	 \end{array}
	\end{equation}
	\begin{equation} \label{q_update_bound_on_bolt_iqueue}
	\arraycolsep=1.1pt\def\arraystretch{1.5}
	\begin{array}{cl}
	 	= &
	 	\displaystyle 
	 	\Big\{
	 	\Big[
	 		Q^{(\text{in})}_{i}(t) + 
	 		\sum_{c' \in p(i)}
			\sum_{i' \in \mathcal{I}_{C}(c')}  
			X_{i', i}(t)
			- \mu_{i}(t)
		\Big]^{+}
	 	\Big\}^{2}
	 	\\ 
	 	\leq & \displaystyle 
	 	\left[ Q^{\text{(in)}}_{i}(t) \right]^2 +
	 	\Big[ 
	 		\sum_{c' \in p(i)}
			\sum_{i' \in \mathcal{I}_{C}(c')}  
			X_{i', i}(t)
	 	\Big]^2 
		+ 
		\left[ \mu_{i}(t) \right]^2
	 	\\
	 	& \displaystyle 
	 	+ 2 Q^{\text{(in)}}_{i}(t)\Big[ 
	 		\sum_{c' \in p(i)}
			\sum_{i' \in \mathcal{I}_{C}(c')}  
				X_{i', i}(t)
			- \mu_{i}(t)
	 	\Big].\nonumber
	 \end{array}
	\end{equation}
	
	3) For bolt instance $i$, regarding its output queue backlog to component $c' \in n(i)$, we have
	\begin{equation}
		\arraycolsep=1.1pt\def\arraystretch{1.3}
		\begin{array}{rl}
			   & \displaystyle 
			   \left[ Q^{\text{(out)}}_{i, c'}(t+1) \right]^2 \\
			   = & \displaystyle 
			   \Big\{ 	
			   	\Big[
					Q^{\text{(out)}}_{i, c'}(t)  
					- 
					\sum_{i' \in \mathcal{I}_{C}(c')}
					X_{i, i'}(t)
					\Big]
					+ \nu_{i, c'}(t)
			   \Big\}^2 
			   \\
			   \leq & \displaystyle 
			   \left[ Q^{\text{(out)}}_{i, \eta}(t) \right]^2 + 
			   \Big[ 
			   		\sum_{i' \in \mathcal{I}_{C}(c')} 
			   		X_{i, i'}(t) 
			   \Big]^2 +
			   \left[ \nu_{i, c'}(t) \right]^2 \\
			   & \displaystyle 
			   + 2 Q^{\text{(out)}}_{i, c'}(t)
			   \Big[ \nu_{i, c'}(t) - 
			   \sum_{i' \in \mathcal{I}_{C}(c')} 
			   	X_{i, i'}(t)
			   \Big].
		\end{array}
	\end{equation}
	
We define $D_{\text{max}}$ as the maximum in-degree and out-degree of components, 
$I_{\text{max}}$ as the maximum number of instances for any component,
and $\gamma_{\text{max}}$ as the maximum transmission capacity in a time slot for one instance.
Also remind the boundedness of all the tuple arrivals, service capacities of instances.
In time $t$,
for spout instance $i$ 
and its succeeding component $c' \in n(i)$,
	\begin{equation}\label{bounds}
	\arraycolsep=1pt\def\arraystretch{1.5}
	    \begin{array}{c}
	        \displaystyle 
	        \lambda_{i, c'}(t + W_{i} + 1) \leq \lambda_{\text{max}},
	    \end{array}
	\end{equation}
	\begin{equation}
		\begin{array}{cll}
	        \displaystyle
	        \sum_{i' \in \mathcal{I}_{C}(c')}
			   	X_{i, i'}(t) &
			\leq & 
			\displaystyle Q^{\text{(out)}}_{i, c'}(t) 
			\leq 
			\sum_{w=0}^{W_{i}}
			\lambda_{i, c'}(t + w) \\
			& \leq & (W_{\text{max}} + 1) \cdot \lambda_{\text{max}}. \\
	    \end{array}
	\end{equation}
For bolt instance $i$, 
regarding its input queue, 
	\begin{equation}
		\mu_{i}(t) \le \mu_{\text{max}}.
	\end{equation}
Since it has at most $D_{\text{max}} \cdot I_{\text{max}}$ preceding instances, and for each $i' \in \mathcal{I}_{C}(c')$ with $c' \in p(i)$,
\begin{equation}
	X_{i', i}(t) \leq
	\sum_{c' \in n(i')}
	\sum_{i'' \in \mathcal{I}_{C}(c')}
		X_{i', i''}(t) \leq D_{\text{max}} I_{\text{max}} \gamma_{\text{max}}.
\end{equation}
Regarding its output queue to component $c' \in n(i)$,
	\begin{equation}
		\nu_{i, c'}(t) \le \nu_{\text{max}},
	\end{equation}
and
	\begin{equation}
		\sum_{i' \in \mathcal{I}_{C}(c')} 
		X_{i, i'}(t) \le \gamma_{i} \le \gamma_{\text{max}}.
	\end{equation}
Therefore, we obtain
\begin{equation}
	\arraycolsep=1pt\def\arraystretch{1.8}
	    \begin{array}{rl}
			& \displaystyle \Delta_V (\boldsymbol{Q}(t))  
	        \\
	        \le & \displaystyle
	        \frac{1}{2} \cdot
	        |\mathcal{I}| \cdot
	        \left[ 
	        	\left(
	        		D_{\text{max}} I_{\text{max}} \gamma_{\text{max}}
	        	\right)^{2} 
	        	+ \left( \mu_{\text{max}} \right)^2
	        \right] 
	        \\
	        + & \displaystyle
	        \frac{\beta}{2} \cdot
	        |\mathcal{I}| \cdot
	        D_{\text{max}} \cdot
	        \left[
	        	\left( W_{\text{max}} + 1 \right)^{2} \cdot
	        	\left( \lambda_{\text{max}} \right)^{2}
	        	+ \left(\lambda_{\text{max}} \right)^{2} 
	        \right]
	        \\
	       \end{array}
	      \end{equation}
\begin{equation}
	\arraycolsep=1pt\def\arraystretch{1.8}
	    \begin{array}{rl}
	        + & \displaystyle
	        \frac{\beta}{2} \cdot
	        |\mathcal{I}| \cdot
	        D_{\text{max}} \cdot
	        \left[
				\left( \nu_{\text{max}} \right)^{2} +
				\left( \gamma_{\text{max}} \right)^{2} 
	        \right]
	        \\
	        + & \displaystyle 
				\mathbb{E}\Big\{
					\sum_{i \in \mathcal{I}}
					\sum_{c' \in p(i)}
					\sum_{i' \in \mathcal{I}_{C}(c')}
					Q^{\text{(in)}}_{i}(t)\left[ 
					X_{i', i}(t) - \mu_{i}(t)
			 		\right]
					\bigg\vert \boldsymbol{Q}(t) \Big\} \\
	       \end{array}
	      \end{equation}
	\begin{equation}
		\arraycolsep=1pt\def\arraystretch{1.8}
			\begin{array}{rl}
			+ & \displaystyle
			\alpha \mathbb{E}\bigg\{
				\sum_{i \in \mathcal{I}}
				\sum_{c' \in n(i)}
				Q^{\text{(out)}}_{i, c'}(t)
			   \bigg[ 
			   	   \lambda_{i, c'}(t + W_{i} + 1)
				   - 
				\\
			& \displaystyle
				   \sum_{i' \in \mathcal{I}_{C}(c')} 
				   	X_{i, i'}(t)
			   \bigg]
			\bigg\vert \boldsymbol{Q}(t) \bigg\} 
		\end{array} \nonumber
\end{equation}
\begin{equation}\label{drift_plus_penalty_expand3}
	\arraycolsep=1pt\def\arraystretch{1.8}
	\begin{array}{rl}
		+ & \displaystyle
			\alpha \mathbb{E}\Big\{
			\sum_{i \in \mathcal{I}}
			\sum_{c' \in n(i)}
			Q^{\text{(out)}}_{i, c'}(t)
		   \Big[ 
		   	   \nu_{i, c'}(t)
			   - 
			   \sum_{i' \in \mathcal{I}_{C}(c')} 
			   	X_{i, i'}(t)
		   \Big]
		   \\
		& \displaystyle
		\left.
		\bigg\vert \boldsymbol{Q}(t) \right\} + \mathbb{E} \left\{ V \Theta(\boldsymbol{X}(t)) \bigg \vert \boldsymbol{Q}(t) \right\},
	    \end{array}
\end{equation}
where $\mathcal{I}^{\text{(spout)}}$ and $\mathcal{I}^{\text{(bolt)}}$ are the set of all spout and bolt instances, respectively. 
Next,
by defining a constant $B$ as
\begin{equation}\label{def_B}
	\arraycolsep=1pt\def\arraystretch{1.8}
	\begin{array}{cl}
		B \triangleq 
		& \displaystyle
	    \frac{1}{2} \cdot
	        |\mathcal{I}| \cdot
	        \left[ 
	        	\left(
	        		D_{\text{max}} I_{\text{max}} \gamma_{\text{max}}
	        	\right)^{2} 
	        	+ \left( \mu_{\text{max}} \right)^2
	        \right] 
	        \\
	        + & \displaystyle
	        \frac{\beta}{2} \cdot
	        |\mathcal{I}| \cdot
	        D_{\text{max}} \cdot
	        \left[
	        	\left( W_{\text{max}} + 1 \right)^{2} \cdot
	        	\left( \lambda_{\text{max}} \right)^{2}
	        	+ \left(\lambda_{\text{max}} \right)^{2} 
	        \right]
	        \\
	        + & \displaystyle
	        \frac{\beta}{2} \cdot
	        |\mathcal{I}| \cdot
	        D_{\text{max}} \cdot
	        \left[
				\left( \nu_{\text{max}} \right)^{2} +
				\left( \gamma_{\text{max}} \right)^{2} 
	        \right].
		\\
	\end{array}
\end{equation}
By substituting (\ref{def_B}) into (\ref{drift_plus_penalty_expand3}) and  
canceling the terms that are irrelevant to the decision variables $\boldsymbol{X}(t)$ (defined as $C(\boldsymbol{Q}(t))$),
then rearranging the term, we obtain
\begin{equation}\label{drift_plus_penalty_expand8}
	    \begin{array}{rl}
	    	\Delta_V (\boldsymbol{Q}(t)) 
	    	& \displaystyle \leq B + C(\boldsymbol{Q}(t)) \\
	        & \displaystyle +
	        \mathbb{E}\Big\{
				\sum_{i \in \mathcal{I}}
				\sum_{c' \in n(i)}
				\sum_{i' \in \mathcal{I}_{C}(c')}
				l_{i, i'}(t)
				X_{i, i'}(t)
			\bigg\vert \boldsymbol{Q}(t) \Big\}, 
	    \end{array}
\end{equation}
in which we define
\begin{equation}
	l_{i, i'}(t) \triangleq
		V \cdot U_{c(i), c(i')}(t)
		+ Q^{\text{(in)}}_{i'}(t)
		- \beta Q^{\text{(out)}}_{i}(t). 
\end{equation}

By minimizing the upper bound of the drift-plus-penalty expression in (\ref{drift_plus_penalty_expand8}), we can minimize the long-term time average of total system cost while stabilizing all processing queues. 
	To transform the minimization of the above bound to minimization of the objective 
	in (\ref{problem-def-small-time-scale}), we have the following statement.
	We denote the objective function at time slot $t$ by $J_{t}(\boldsymbol{X})$ with decision $\boldsymbol{X}$, and its optimal solutions by $\boldsymbol{X}^{*}$. 
	Hence, for any other feasible scheduling decisions  $\boldsymbol{X}$ made during time slot $t$, we have 
	\begin{equation}\label{inequality_J}
		J_t(\boldsymbol{X}) \ge J_t(\boldsymbol{X}^{*})	
	\end{equation}
	By taking the conditional expectation on both sides conditional on $\boldsymbol{Q}(t)$, we have
	\begin{equation}\label{inequality_J_expectation}
		\mathbb{E} \left\{  
			J_t(\boldsymbol{X})
		\bigg\vert \boldsymbol{Q}(t) \right\}
		\ge 
		\mathbb{E} \left\{  	
		J_t(\boldsymbol{X}^{*})	
		\bigg\vert \boldsymbol{Q}(t) \right\},
	\end{equation}
	for any feasible $\boldsymbol{X}$.
	Inequality (\ref{inequality_J_expectation}) reveals that $\boldsymbol{X}^{*}$ minimizes the conditional expectation of $J_t(\boldsymbol{X})$, thusly minimizing the upper bound of drift-plus-penalty in (\ref{drift_plus_penalty_expand8}). 
	In such a way, instead of directly solving the long-term stochastic optimization problem (\ref{problem-def-small-time-scale}), we can opportunistically choose a feasible association to solve the following problem in each time slot.
	\begin{equation}\label{problem_per_time_slot}
	\setstretch{1.3}
	\begin{array}{cl}
	    \underset{\boldsymbol{X}}{\text{Minimize}} & J_t(\boldsymbol{X}) \\
        \text{Subject to} &
		\displaystyle
		0 \le 
		\sum_{c' \in n(i)} \sum_{i' \in \mathcal{I}_{C}(c')} X_{i, i'} 
		\le \gamma_{i}, \ \forall\ i \in \mathcal{I}, \\
		& \displaystyle
		\sum_{i' \in \mathcal{I}_{C}(c')} X_{i, i'} \le Q_{i, c'}^{\text{(out)}}(t), \  
		\forall\ i \text{ and } c' \in n(i).
	\end{array}
	\end{equation}
	\IEEEQED

\section{Proof of Theorem 1}
We assume there exists an S-only algorithm \cite{neely2010stochastic} that achieves the optimal time-average total cost (infimum) $\Theta^*$ with action $\tilde{\boldsymbol{X}}(t)$ for $t= \{0,1,2,\cdots \}$. 
	From \cite{neely2010stochastic}, we know that to ensure queue stability, the expectation of arrival rate must be no more than the expectation of service rate, with a difference of $\epsilon \geq 0 $.
	Denote $\boldsymbol{X}'(t)$ as the decisions over time, and $\Theta'(t)$ as the corresponding communication cost given by POTUS. 
	The one-slot drift-plus-penalty is
	\begin{equation}\label{proof_step2}
	    \begin{array}{l}
	    	\displaystyle
	        \mathbb{E} \left\{ L(\mathbf{Q}(t+1)) - L(\mathbf{Q}(t)) \big\vert \mathbf{Q}(t)\right\} + V \mathbb{E} \left\{ \Theta'(t) \big\vert \mathbf{Q}(t)\right\}\\
	        \displaystyle
	        \leq B + V \mathbb{E} \left\{ \Theta^*(t) \big\vert \mathbf{Q}(t)\right\} - \epsilon \mathbb{E} \left\{ h(t) \big\vert \mathbf{Q}(t)\right\}.
	    \end{array}
	\end{equation}
	Taking expectation over both sides, we obtain
	\begin{equation}\label{proof_step3}
	    \begin{split}
	        \displaystyle &\mathbb{E} \left\{ L(\mathbf{Q}(t+1))\right\} - \mathbb{E}\left\{L(\mathbf{Q}(t))\right\} + V\mathbb{E} \left\{\Theta'(t)\right\}\\ 
	        \leq &~ B + V \mathbb{E} \left\{ \Theta^*(t) \right\} - \epsilon \mathbb{E} \left\{ h(t)\right\}. \\
	    \end{split}
	\end{equation}
	Summing over $t=\{0, 1, \cdots, T-1\}$, we have 
	\begin{equation}\label{proof_step4}
	    \begin{split}
	        \mathbb{E} &\left\{ L(\mathbf{Q}(T-1)) \right\} - \mathbb{E} \left\{ L(\mathbf{Q}(0))\right\} + V \mathbb{E} \Big\{ \sum_{t=0}^{T-1} \Theta'(t) \Big\} \\
	        \leq &~ BT + V \mathbb{E} \Big\{\sum_{t=0}^{T-1} \Theta^*(t) \Big\} - \epsilon \mathbb{E} \left\{ h(t) \right\}.
	    \end{split}
	\end{equation}
	By (\ref{proof_step4}), we are ready to show the two performance bounds in \textit{Theorem 1}.
	
	1) Dividing both sides of (\ref{proof_step4}) by $VT$, rearranging items and neglecting the non-positive on the right side, we obtain
	
	\begin{equation}\label{performance_cost1}
	    \begin{split}
	    \frac{1}{T}\sum_{t=0}^{T-1} \mathbb{E} \left\{ \Theta'(t) \right\}
	    \leq \frac{B}{V} + \frac{\mathbb{E}\{L(\boldsymbol{Q}(0))\}}{VT} + \frac{1}{T}\sum_{t=0}^{T-1} \mathbb{E} \left\{ \Theta^*(t) \right\}.
	    \end{split}
	\end{equation}
	As $T\rightarrow \infty$, we have
	\begin{equation}\label{performance_cost2}
	    \begin{split}
	    \limsup_{T\rightarrow \infty} \frac{1}{T}&\sum_{t=0}^{T-1} \mathbb{E} \left\{ \left( \Theta'(t) \right) \right\} \\
	    &\leq \limsup_{T\rightarrow \infty} \frac{1}{T}\sum_{t=0}^{T-1} \mathbb{E} \left\{ \left( \Theta^*(t) \right) \right\} + \frac{B}{V} \\ 
	    \end{split}.
	\end{equation}
	
	2) Similarly, dividing both sides of (\ref{proof_step4}) by $\epsilon T$, we obtain
	\begin{equation}\label{performance_queue1}
	    \begin{split}
	    &\frac{1}{T}\sum_{t=0}^{T-1} \mathbb{E} \left\{ h(t) \right\} \\
	    &\leq \frac{B}{\epsilon} + \frac{\mathbb{E}\{L(\mathbf{Q}(0))\}}{\epsilon T} + \frac{V\sum_{t=0}^{T-1}\mathbb{E} \left\{ \Theta^*(t) \right\}}{\epsilon T}. \\
	    \end{split}
	\end{equation}
	As $T\rightarrow \infty$, we obtain
	\begin{equation}\label{performance_queue3}
		\limsup_{T\rightarrow \infty} \frac{1}{T}\sum_{t=0}^{T-1} \mathbb{E} \left\{ h(t) \right\} \leq \frac{V \bar{\Theta}^{*}}{\epsilon} + \frac{B}{\epsilon}.
	\end{equation}
	\IEEEQED

\end{document}